\documentclass[%
 reprint,
 amsmath,amssymb,
 aps,
]{revtex4-1}

\usepackage{color}
\usepackage{graphicx}
\usepackage{dcolumn}
\usepackage{bm}

\begin{document}

\preprint{APS/123-QED}

\title{Quark-Pauli effects in three octet-baryons}

\author{C. Nakamoto$^1$}
\author{Y. Suzuki$^{2,\,3}$}%
\affiliation{%
$^1$National Institute of Technology, Suzuka College, Suzuka 510-0294, Japan\\
$^2$Department of Physics, Niigata University, Niigata 950-2181, Japan\\
$^3$RIKEN Nishina Center, Wako 351-0198, Japan
}%




\date{\today}

\begin{abstract}
To sustain a neutron star with about two times the solar mass, 
multi baryons including hyperons are expected to produce 
repulsive effects in the interior of its high baryon-density region. 
To examine possible quark-Pauli repulsion among the baryons,  
we solve the eigenvalue problem of 
the quark antisymmetrizer for three octet-baryons  
that are described by most compact spatial configurations. 
We find that 
the Pauli blocking effect is weak in the $\Lambda nn$ system, 
while it is strong in the $\Sigma^-nn$ system. The appearance of 
the $\Sigma^-$ hyperon is suppressed in the neutron star interior but 
no quark-Pauli repulsion effectively works for the $\Lambda$ hyperon. 
\end{abstract}

\maketitle


\section{\label{sec:level1}INTRODUCTION}

Recently, the properties of multi baryons including hyperons ($Y$'s) 
have attracted much attention 
in the study of neutron stars. 
Though the neutron star is primarily composed of neutrons ($n$'s), 
the presence of the $Y$ appears to be energetically unavoidable \cite{Soviet1960}. 
If the $n$ becomes superabundant in the interior of the neutron star 
and the neutron Fermi energy greatly increases, 
the $n$ becomes unstable against decaying into the $\Lambda$ hyperon via the weak interaction.  
On top of that, if the electron ($e^-$) chemical potential grows 
with increase of the baryon density in the neutron star, the 
$\Sigma^-$ hyperon may be formed through the weak interaction, $n+e^-\rightarrow\Sigma^-+\nu$. Furthermore, it is suggested that the $\Sigma^-$ may appear at a lower density earlier than the $\Lambda$ in spite of the fact that the $\Sigma^-$ is more massive than the $\Lambda$~\cite{Bombaci2005}. 
It is also suggested that the $\Xi^-$ hyperon may appear at a relatively low density depending on the strength of $\Xi^-$ attraction in the interior of the neutron star \cite{PRC89_2014, Miyatsu2015}. 

The appearance of $Y$'s in the neutron star, however, leads to a softening 
of the equation of state~\cite{LNP2001}. Because of this softening 
the maximum mass of the neutron star predicted by solving the equation of 
state with $Y$-nucleon ($N$) and $YY$ interactions used 
in the hypernuclear physics 
turns out to be incompatible with the recent observation~\cite{2SM2010,2SM2013} that finds the neutron star with about double solar mass. 
A resolution of this problem calls for a mechanism that could provide additional repulsion 
to make the equation of state stiffer \cite{Vidana2013}.  
As a candidate of the additional repulsion, various proposals have been made, 
e.g., the vector-meson exchange in the baryon-baryon interactions~\cite{vector2012,OertelJPG2015}, 
the repulsive $\Lambda nn$ three-body force \cite{Lonardoni2015}, 
a universal three-baryon repulsion \cite{Takatsuka2008, Yamamoto2014}, 
the cold quark-matter rather than the hadronic matter~\cite{Fraga2015}.  
See also Refs.~\cite{BombaciHYP2015,vidana2016} for other cases. 


Both of $N$'s and $Y$'s are members of octet-baryons ($B_8$'s). 
Describing them as three-quark clusters,  
we investigate quark-Pauli effect in the three-$B_8$ systems 
because it could be responsible for the needed additional repulsion. 
The quark-Pauli effect becomes most apparent when the three 
baryons strongly overlap. Any three-$B_8$ channel that is (almost) 
Pauli-forbidden provides such short-ranged three-body repulsion 
that is independent of the baryon-baryon interactions. 

The quark-Pauli effect in the two-$B_8$ system has already been studied \cite{OSY1987,central1995}. The effect  
often leads to important repulsion regardless of the detail of the baryon-baryon interaction~\cite{central1997}. 
For example, the repulsive $\Sigma$ single-particle potential 
in nuclei~\cite{Noumi2002} is considered 
to originate from the strong Pauli repulsion in the $\Sigma N(I=\frac{3}{2})\,^3S_1$ state \cite{Kohno2000}.
The most recent Nijmegen ESC08 potential incorporates this 
quark effect phenomenologically 
because it is difficult to achieve strongly repulsive short-range 
interactions in this channel \cite{ESC08}. 
There are some earlier studies on the quark-Pauli effect in the three- and more-baryon systems. See, for example, Refs.~\cite{TSH1982,SHT1982,TS1986,suzuki84,Maltman85}. These are mostly for multi-baryon systems composed of $N$'s and 
a single hyperon such as $N^n$ and $\Lambda N^n$.  

The plan of this paper is as follows. 
We construct antisymmetric three-$B_8$ states 
in sec.~\ref{sec:level2} with a particular 
emphasis on the most compact spatial quark configurations. We discuss 
in sec.~\ref{sec:level3} the quark-Pauli effect by solving the eigenvalue 
problem of the antisymmetrizer of 9 quarks. Conclusion is drawn in 
sec.~\ref{sec:level4}.

\section{\label{sec:level2}THREE OCTET-BARYON STATES}

The octet baryons ($B_8$) with spin $S=\frac{1}{2}$ include $N, \Lambda, \Sigma$, and $\Xi$, all belonging to a member of the flavor SU(3) symmetry $(\lambda \mu)=(11)$. We use the Elliott notation for the SU(3) group~\cite{Elliott}. 
The $B_8$'s are classified by the SU(2)$\times$U(1) subgroup label, $a=YI$, the hypercharge $Y$ and the isospin $I$: $N(YI=1\frac{1}{2})$, $\Lambda(00)$, $\Sigma(01)$, $\Xi(-1\, \frac{1}{2})$. 
Assuming that the $B_8$ is a three-quark cluster, we describe its orbital 
part $\phi^{(\rm orb)}(123)$ by the $(0s)^3$ harmonic-oscillator wave function with a common size parameter.
Since $\phi^{(\rm orb)}(123)$ is totally symmetric and 
the $B_8$ color wave function $C(123)$ is totally antisymmetric, its spin-flavor part represented by $W^{[3]}(123)$ must be totally symmetric, as indicated by [3] symmetry. By specifying the $z$-components of the 
spin and the isospin by $S_z$ and $I_z$, respectively, a full quark-model description of $B_8$ reads \cite{fss2}
\begin{align}
\psi_{(11)aS_z I_z}(123)=\phi^{(\rm orb)}(123)W^{[3]}_{a S_z I_z}(123)C(123).
\label{wf.baryon}
\end{align}
More explicitly, $W^{[3]}_{a S_z I_z}(123)$ is given by 
\begin{align}
&W^{[3]}_{ a S_z I_z}(123)\notag \\
&=\frac{1}{\sqrt{2}}\sum_{S'(\lambda' \mu')=0(01), 1(20)} 
\big[[w_{\frac{1}{2}}(1)w_{\frac{1}{2}}(2)]_{S'} w_{\frac{1}{2}}(3)\big]_{\frac{1}{2} S_z}\notag \\
&\qquad \qquad \times 
\big[[F_{(10)}(1)F_{(10)}(2)]_{(\lambda' \mu')}F_{(10)}(3)\big]_{(11) a I_z},
\end{align}
where $w_{\frac{1}{2}}$ and $F_{(10)}$ are the spin and flavor functions of the single quark. The square bracket $[\ \ \ ]$ is used to stand for spin and/or 
flavor SU(3) couplings. 

Equation~(\ref{wf.baryon}) gives the normalized $B_8$ wave function that satisfies 
the required symmetry at the quark level. By combining two $B_8$ wave functions, it is possible 
to express the spin-isospin coupled basis in terms of a combination of the spin-flavor coupled basis~\cite{central1995,fss2NN}. Physically 
allowed two-baryon states have to satisfy the generalized Pauli principle that demands the total wave function to be antisymmetric under the exchange of quarks. We extend this to a special three-$B_8$ state in which all nine quarks occupy the same $0s$ harmonic-oscillator function. The orbital configuration of that state is most compact and such three-$B_8$ state is expected to be most 
strongly influenced by the quark-Pauli principle. 

To construct the fully antisymmetric 9-quark states, we first start from 
the three-$B_8$ configuration that is antisymmetric under the exchange of 
baryons.  
The $(0s)^9$ configuration is apparently symmetric under the exchange of 
baryons. The color part 
is also totally symmetric with respect to the exchange of baryons. Therefore 
the spin-flavor part of the 
three-$B_8$ state must be antisymmetric under the exchange of baryons. 
To construct such three-$B_8$ spin-flavor states, we combine   
the spin-isospin coupled two-$B_8$ state, $[W^{[3]}_{a_1}(123) W^{[3]}_{a_2}(456)]_{S' a'}$, with the third $B_8$ as follows:
\begin{align}
\big[[ W^{[3]}_{a_1}(123) W^{[3]}_{a_2}(456) ]_{S' a'} W^{[3]}_{a_3}(789) \big]_{S a S_z I_z}. 
\label{SFSI}
\end{align}
Here $S$ is the total spin that couples $S'$ with $\frac{1}{2}$ and $a$ includes the isospin coupling of $I'$ and $I_3$. 
The hypercharge is trivially given as $Y'=Y_1+Y_2$ and $Y=Y'+Y_3$. 
The $z$ components of $S$ and $I$, $S_z$ and $I_z$, are abbreviated in what 
follows throughout this section. 
Since in this section we focus on constructing such three-$B_8$ 
spin-flavor functions that are antisymmetric under the baryon exchange, we 
suppress the quark labels and simplify 
$W^{[3]}_{a}(123)$ by $B_{a}(1)$ and express Eq.~(\ref{SFSI}) as  
\begin{align}
\big[[B_{a_1}(1)B_{a_2}(2)]_{S'a'}B_{a_3}(3)\big]_{Sa}.
\label{base}
\end{align}
The fully antisymmetrized spin-flavor function constructed from Eq.~(\ref{base}) is 
\begin{align}
&\Psi^{(\rm SF)}_{Sa}(a_1a_2a_3,S'a') \notag \\
&={\cal N} \Big\{\ \big[[B_{ a_1}(1)B_{ a_2}(2)]_{S'a'}B_{ a_3}(3)\big]_{Sa} \notag \\
&\quad \quad -\big[[B_{a_1}(2)B_{a_2}(1)]_{S'a'}B_{a_3}(3)\big]_{Sa} \notag \\
&\quad \quad +\big[[B_{a_1}(2)B_{a_2}(3)]_{S'a'}B_{a_3}(1)\big]_{Sa} \notag \\
&\quad \quad -\big[[B_{a_1}(3)B_{a_2}(2)]_{S'a'}B_{a_3}(1)\big]_{Sa} \notag \\
&\quad \quad +\big[[B_{a_1}(3)B_{a_2}(1)]_{S'a'}B_{a_3}(2)\big]_{Sa} \notag \\
&\quad \quad -\big[[B_{a_1}(1)B_{a_2}(3)]_{S'a'}B_{a_3}(2)\big]_{Sa} \Big\}\,,
\label{antisym}
\end{align}
which is characterized by $S, a$ and $S', a'$ as well as $a_1, a_2, a_3$. Here ${\cal N}$ is a normalization constant.

With use of the angular-momentum recoupling or Racah coefficients $U$ in unitary form, the function~(\ref{antisym}) can be expressed as 
\begin{widetext}
\begin{align}
&\Psi^{(\rm SF)}_{Sa}(a_1a_2a_3,S'a')\notag \\
&={\cal N}  \Bigg\{ 
[[B_{a_1}(1)B_{a_2}(2)]_{S'a'}B_{a_3}(3)]_{Sa} +(-1)^{S'+I_1+I_2-I'} [[B_{a_2}(1)B_{a_1}(2)]_{S'a'}B_{a_3}(3)]_{Sa} \notag \\
&\ \ +\sum_{S_{12}I_{12}}\Big[
(-1)^{1/2+S_{12}-S+I_1+I_{12}-I} U(\textstyle{\frac{1}{2}} \frac{1}{2} S \frac{1}{2}; S'S_{12}) U(I_1 I_2 I I_3;I'I_{12}) [[B_{a_2}(1)B_{a_3}(2)]_{S_{12}a_{12}}B_{a_1}(3)]_{Sa} \notag \\
&\qquad +(-1)^{1/2-S+I_1+I_2+I_3-I} U(\textstyle{\frac{1}{2}} \frac{1}{2} S \frac{1}{2};S_{12}S') U(I_3 I_2 I I_1;I_{12}I') [[B_{a_3}(1)B_{a_2}(2)]_{S_{12}a_{12}}B_{a_1}(3)]_{Sa} \notag \\
& \qquad +(-1)^{1/2+S'-S+I'+I_3-I} U(\textstyle{\frac{1}{2}} \frac{1}{2} S \frac{1}{2};S_{12}S') U(I_3 I_1 I I_2;I_{12}I') [[B_{a_3}(1)B_{a_1}(2)]_{S_{12}a_{12}}B_{a_2}(3)]_{Sa} \notag \\
&\qquad +(-1)^{1/2-S'+S_{12}-S+I_1+2I_2+I_{12}-I'-I} U(\textstyle{\frac{1}{2}} \frac{1}{2} S \frac{1}{2};S'S_{12}) U(I_2 I_1 I I_3;I'I_{12}) [[B_{a_1}(1)B_{a_3}(2)]_{S_{12}a_{12}}B_{a_2}(3)]_{Sa}
\Big]\Bigg\},
\label{formula3}
\end{align}
\end{widetext}
where the order of particle labels is always arranged to 1, 2 and 3, while that of the baryon species, $a_1a_2a_3$, is changed appropriately. 
In what follows, 
we often abbreviate $[[B_{a_1}(1)B_{a_2}(2)]_{S'a'}B_{a_3}(3)]_{Sa}$ as $[[B_{a_1}B_{a_2}]_{S'a'}B_{a_3}]_{Sa}$. 

It should be noted that, for a given set of $a_1a_2a_3$, 
the functions~(\ref{formula3}) generated using all possible values of $S'a'$ 
provide a full set of the 
antisymmetric functions but they are not always independent. Also the order of 
$a_1a_2a_3$ is not important.    
For example, two independent functions, 
$\Psi^{(\rm SF)}_{\frac{1}{2}\, 1\frac{1}{2}}(\Sigma \Lambda N, 01)$ and 
$\Psi^{(\rm SF)}_{\frac{1}{2}\, 1\frac{1}{2}}(\Sigma \Lambda N, 11)$, are 
related to  
$\Psi^{(\rm SF)}_{\frac{1}{2}\, 1\frac{1}{2}}(\Lambda N \Sigma, 0\frac{1}{2})$ and 
$\Psi^{(\rm SF)}_{\frac{1}{2}\, 1\frac{1}{2}}(\Lambda N \Sigma, 1\frac{1}{2})$ as 
\begin{align}
&\Psi^{(\rm SF)}_{\frac{1}{2}\, 1\frac{1}{2}}(\Sigma \Lambda N, 01) \notag \\
&=\textstyle{\frac{1}{2}} \Psi^{(\rm SF)}_{\frac{1}{2}\, 1\frac{1}{2}}(\Lambda N \Sigma, 0\textstyle{\frac{1}{2}}) 
+\frac{\sqrt{3}}{2} \Psi^{(\rm SF)}_{\frac{1}{2}\, 1\frac{1}{2}}(\Lambda N \Sigma, 1\frac{1}{2}),\notag \\
&\Psi^{(\rm SF)}_{\frac{1}{2}\, 1\frac{1}{2}}(\Sigma \Lambda N, 11) \notag \\
&=-\textstyle{\frac{\sqrt{3}}{2}} \Psi^{(\rm SF)}_{\frac{1}{2}\, 1\frac{1}{2}}(\Lambda N \Sigma, 0\textstyle{\frac{1}{2}}) 
+\frac{1}{2}\Psi^{(\rm SF)}_{\frac{1}{2}\, 1\frac{1}{2}}(\Lambda N \Sigma, 1\frac{1}{2}).
\end{align}
In the case of $\Sigma NN$ with $S=\frac{1}{2}, I=1$, there are four possible $S'a'$ values but only two independent, antisymmeytic functions can be 
constructed. By introducing a label $v$ to enumerate the antisymmetric, orthogonal 
functions, they read 
\begin{align}
&\Psi^{(\rm SF)}_{\frac{1}{2}\, 21}(\Sigma NN,v=1) \notag \\
& \ \ =\textstyle{\frac{1}{\sqrt{3}}}[[NN]_{021}\Sigma]_{\frac{1}{2}\,2 1} +\textstyle{\frac{1}{3}}
[[\Sigma N]_{01 \frac{1}{2}}N]_{\frac{1}{2}\,2 1} \notag \\ 
&\ \ \ +\textstyle{\frac{1}{\sqrt{18}}}[[\Sigma N]_{01\frac{3}{2}}N]_{\frac{1}{2}\, 2 1} -\textstyle{\frac{1}{\sqrt{3}}}[[\Sigma N]_{11\frac{1}{2}}N]_{\frac{1}{2}\, 2 1}  \notag \\
& \ \ \ -\textstyle{\frac{1}{\sqrt{6}}}[[\Sigma N]_{11\frac{3}{2}}N]_{\frac{1}{2}\, 2 1},\nonumber \\ 
&\Psi^{(\rm SF)}_{\frac{1}{2}\, 21}(\Sigma NN, v=2) \notag \\
& \ \ =\textstyle{\frac{1}{\sqrt{3}}}[[NN]_{120}\Sigma]_{\frac{1}{2}\, 21} +\textstyle{\frac{1}{\sqrt{6}}}[[\Sigma N]_{01 \frac{1}{2}}N]_{\frac{1}{2}\, 21} \notag \\
&\ \ \ -\textstyle{\frac{1}{\sqrt{3}}}[[\Sigma N]_{01\frac{3}{2}}N]_{\frac{1}{2}\, 21} +\textstyle{\frac{1}{\sqrt{18}}}[[\Sigma N]_{11\frac{1}{2}}N]_{\frac{1}{2}\, 21} \notag \\
&\ \ \ -\textstyle{\frac{1}{3}}[[\Sigma N]_{11\frac{3}{2}}N]_{\frac{1}{2}\, 21}. 
\end{align}
In the case where only one antisymmetric function is possible, the label $v$ is 
suppressed. 
All of the totally antisymmetric spin-flavor functions of three-$B_8$ systems 
are tabulated for both $S=\frac{1}{2}$ and $\frac{3}{2}$ in Appendix A of 
Supplemental Material~\cite{SM}. 
The spin part of the three-$B_8$ state with $S=\frac{3}{2}$ is totally 
symmetric under the baryon exchange, 
so that its flavor part is totally antisymmetric.

\begin{table}
\caption{Short-hand notations $|\lambda \mu \rangle_n$ 
for the antisymmetric spin-flavor functions in the flavor SU(3) basis, 
$|S_{12}(\lambda_{12}\mu_{12}){\rho_{12}};S(\lambda\mu)\rho a \rangle$, where  
the multiplicity label $\rho$ is explicitly written for the 
SU(3) couplings of $(11)\times(11)\rightarrow(11)$ 
and $(22)\times(11)\rightarrow(22)$, but it is suppressed in 
multiplicity-free cases. 
The total spin $S$ and the label $a$ are abbreviated in 
the short-hand notation.}
\label{short_sf}
\begin{center}
{\renewcommand\arraystretch{1.7}
\begin{minipage}{40mm}
\begin{tabular}{|c|c|} \hline
$|41\rangle_1$ & $\left|\,0(22)\,;\,S(41)a\,\right\rangle$ \\
$|41\rangle_2$ & $\left|\,1(30)\,;\,S(41)a\,\right\rangle$ \\
 \hline
$|30\rangle_1$ & $\left|\,0(22)\,;\,S(30)a\,\right\rangle$ \\
$|30\rangle_2$ & $\left|\,1(30)\,;\,S(30)a\,\right\rangle$ \\
$|30\rangle_3$ & $\left|\,1(11){1}\,;\,S(30)a\,\right\rangle$ \\
$|30\rangle_4$ & $\left|\,0(11){2}\,;\,S(30)a\,\right\rangle$ \\
 \hline
$|22\rangle_1$ & $\left|\,0(22)\,;\,S(22){1}\,a\,\right\rangle$ \\
$|22\rangle_2$ & $\left|\,0(22)\,;\,S(22){2}\,a\,\right\rangle$ \\
$|22\rangle_3$ & $\left|\,1(30)\,;\,S(22)a\,\right\rangle$ \\
$|22\rangle_4$ & $\left|\,1(03)\,;\,S(22)a\,\right\rangle$ \\
$|22\rangle_5$ & $\left|\,1(11){1}\,;\,S(22)a\,\right\rangle$ \\
$|22\rangle_6$ & $\left|\,0(11){2}\,;\,S(22)a\,\right\rangle$ \\
 \hline
$|14\rangle_1$ & $\left|\,0(22)\,;\,S(14)a\,\right\rangle$ \\
$|14\rangle_2$ & $\left|\,1(03)\,;\,S(14)a\,\right\rangle$ \\
 \hline
\end{tabular}
\end{minipage}
\begin{minipage}{40mm}
\begin{tabular}{|c|c|} \hline
$|11\rangle_1$ & $\left|\,0(22)\,;\,S(11)a\,\right\rangle$ \\
$|11\rangle_2$ & $\left|\,1(30)\,;\,S(11)a\,\right\rangle$ \\
$|11\rangle_3$ & $\left|\,1(03)\,;\,S(11)a\,\right\rangle$ \\
$|11\rangle_4$ & $\left|\,1(11){1}\,;\,S(11){1}\,a\,\right\rangle$ \\
$|11\rangle_5$ & $\left|\,1(11){1}\,;\,S(11){2}\,a\,\right\rangle$ \\
$|11\rangle_6$ & $\left|\,0(11){2}\,;\,S(11){1}\,a\,\right\rangle$ \\
$|11\rangle_7$ & $\left|\,0(11){2}\,;\,S(11){2}\,a\,\right\rangle$ \\
$|11\rangle_8$ & $\left|\,0(00)\,;\,S(11)a\,\right\rangle$ \\
 \hline
$|03\rangle_1$ & $\left|\,0(22)\,;\,S(03)a\,\right\rangle$ \\
$|03\rangle_2$ & $\left|\,1(03)\,;\,S(03)a\,\right\rangle$ \\
$|03\rangle_3$ & $\left|\,1(11){1}\,;\,S(03)a\,\right\rangle$ \\
$|03\rangle_4$ & $\left|\,0(11){2}\,;\,S(03)a\,\right\rangle$ \\
 \hline
$|00\rangle_1$ & $\left|\,0(11){1}\,;\,S(00)a\,\right\rangle$ \\
$|00\rangle_2$ & $\left|\,0(11){2}\,;\,S(00)a\,\right\rangle$ \\
\hline
\end{tabular}
\end{minipage}}
\end{center}
\label{short-sfwf}
\end{table}

The flavor SU(3) symmetry was used in advantage in studying 
the $B_8 B_8$ interaction in Ref.~\cite{fss2}. This is based on the 
assumption that the underlying Hamiltonian 
for the octet-baryon system is approximately SU(3)-scalar. 
Similarly we expect that the flavor SU(3) symmetry plays an important 
role in the three-$B_8$ systems. To exploit this possibility, we 
represent all three-$B_8$ states obtained in the  spin-isospin 
coupled basis in the flavor SU(3) basis. 
This is done in exactly the same way as 
the two-baryon case~\cite{central1995,fss2NN} with use of 
reduced SU(3) Wigner coefficients~\cite{DraayerAkiyama}:    
\begin{align}
&[B_{a_1}(1) B_{a_2}(2)]_{S_{12} a_{12}} \notag \\
& \ =\sum_{(\lambda_{12}\mu_{12})\rho_{12}}
\langle (11)a_1\, (11)a_2 || (\lambda_{12} \mu_{12}) a_{12} \rangle_{\rho_{12}} \notag \\
& \quad \times [B_{(11)}(1) B_{(11)}(2)]_{S_{12} (\lambda_{12}\mu_{12})\rho_{12}a_{12}},
\label{SU3_coupling}
\end{align}
where $(\lambda_{12} \mu_{12})$ takes (22), (11), (00) for $S_{12}=0$ and 
(30), (03), (11) for $S_{12}=1$, respectively. 
The label $\rho_{12}$ distinguishes 
possible multiple occurrences of $(\lambda_{12} \mu_{12})$. Two representations appear for $(\lambda_{12} \mu_{12})=(11)$, and  
$\rho_{12}=1$ stands for an antisymmetric coupling, while $\rho_{12}=2$ a symmetric coupling.
Further application of the SU(3) coupling with $B_{a_3}(3)$ makes it possible to express 
the three-$B_8$ spin-isospin coupled state as follows: 
\begin{align}
&[[B_{a_1}(1) B_{a_2}(2)]_{S_{12} a_{12}}B_{a_3}(3)]_{Sa} \notag \\
&=\sum_{(\lambda_{12}\mu_{12})\rho_{12}(\lambda\mu)\rho}
\langle (11)a_1\, (11)a_2 || (\lambda_{12} \mu_{12}) a_{12} \rangle_{\rho_{12}} \notag \\
&\quad\quad\quad\quad\quad\quad\times\langle (\lambda_{12}\mu_{12})a_{12}\, (11)a_3 || (\lambda\mu)a \rangle_\rho \notag \\
&\quad\quad\quad\quad\quad\quad\times|S_{12}(\lambda_{12}\mu_{12}){\rho_{12}};S(\lambda\mu)\rho a\rangle
\label{SU3_coupling2}
\end{align}
with
\begin{align}
&|S_{12}(\lambda_{12}\mu_{12}){\rho_{12}};S(\lambda \mu)\rho a\rangle \notag \\
& = 
[[B_{(11)}(1) B_{(11)}(2) ]_{S_{12}(\lambda_{12}\mu_{12})\rho_{12}} B_{(11)}(3)]_{S (\lambda \mu)\rho a}.
\label{table1}
\end{align} 
For the sake of convenience, short-hand notations for 
$|S_{12}(\lambda_{12}\mu_{12}){\rho_{12}};S(\lambda\mu)\rho a \rangle$ 
are introduced as shown in Table \ref{short_sf}.

Applying Eq.~(\ref{SU3_coupling2}) in Eq.~(\ref{formula3}) and following the 
construction of the function $\Psi_{Sa}(a_1a_2a_3,v)$ 
defines the totally antisymmetric spin-flavor three-$B_8$ state 
in the flavor SU(3) basis as follows:
\begin{align}
\Psi^{(\rm SF)}_{Sa}&(a_1a_2a_3,v) =\sum_{S_{12}(\lambda_{12}\mu_{12})\rho_{12}(\lambda \mu)\rho} \notag \\
&  G(a_1a_2a_3, v, S_{12}(\lambda_{12}\mu_{12})\rho_{12}, (\lambda \mu)\rho ; Sa) \notag \\
& \times |S_{12}(\lambda_{12}\mu_{12}){\rho_{12}} ; S(\lambda \mu)\rho a\rangle.
\label{3B.SF}
\end{align}
Tables~\ref{su3_half} and \ref{su3_three-half} 
tabulate the coefficients $G$ for some interesting three-$B_8$ systems 
including 
(i) $NNN$,  (ii) $YNN$ and $YYN$ that couple each other, (iii)
high-isospin systems that may be important in the neutron-star interior,   
and (iv) those systems that lead to almost Pauli-forbidden states.
Other three-$B_8$ systems are tabulated in Appendix B of Supplemental Material~\cite{SM}. 

As Table \ref{su3_half} shows for the $S=\frac{1}{2}$ case, 
a group of $NNN\, (I=\frac{1}{2})$, 
$\Xi\Sigma\Sigma\, (I=\frac{5}{2})$, and $\Xi\Xi\Sigma\, (I=2)$ states 
belongs to $|14\rangle_1-|14\rangle_2$, and likewise a group of 
$\Sigma NN\, (I=2)$, $\Sigma\Sigma N\, (I=\frac{5}{2})$, 
and $\Xi\Xi\Xi\, (I=\frac{1}{2})$ belongs to   
$|41\rangle_1-|41\rangle_2$. In the $S=\frac{3}{2}$ sector, 
Table~\ref{su3_three-half} shows that  
$\Sigma NN\, (I=1)$, $\Xi\Sigma N\, (I=2)$, and 
$\Xi\Xi\Sigma\, (I=1)$ systems are all expressed by $2|22\rangle_3-2|22\rangle_4-|22\rangle_5$.

\begin{table*}
\caption{\label{su3_half}
Coefficients $G$ in Eq. (\ref{3B.SF}) for 
some three-$B_8$ systems with $S=1/2$. 
The label $v$ distinguishes the multiple occurrence of the 
orthogonal, antisymmetric states for a given $SYI$.  
This table continues.
}
\begin{ruledtabular}
\begin{tabular}{ccccccccccccc}
$Y\,I$ & $3\,\frac{1}{2}$ & \multicolumn{2}{c}{2\,0} & \multicolumn{3}{c}{2\,1} & 2\,2 & \multicolumn{5}{c}{$1\,\frac{1}{2}$} \\
\cline{1-1}\cline{2-2}\cline{3-4}\cline{5-7}\cline{8-8}\cline{9-13}
 & $NNN$ & $\Lambda NN$ & $\Sigma NN$ & $\Lambda NN$ & $\Sigma NN$
 & $\Sigma NN$ & $\Sigma NN$ & $\Xi NN$ & $\Xi NN$ & $\Lambda\Lambda N$ & $\Sigma\Sigma N$ & $\Sigma\Sigma N$ \\ 
 & & & & & v=1 & v=2 & & v=1 & v=2 & & v=1 & v=2 \\
\cline{1-1}\cline{2-2}\cline{3-4}\cline{5-7}\cline{8-8}\cline{9-13}
$|41\rangle_1$ &  &  &  &  &  &  & $-\frac{1}{\sqrt{2}}$ &  &  &  &  &  \\
$|41\rangle_2$ &  &  &  &  &  &  & $\frac{1}{\sqrt{2}}$ &  &  &  &  &  \\
$|30\rangle_1$ &  &  &  &  &  &  &  &  &  &  &  &  \\
$|30\rangle_2$ &  &  &  &  &  &  &  &  &  &  &  &  \\
$|30\rangle_3$ &  &  &  &  &  &  &  &  &  &  &  &  \\
$|30\rangle_4$ &  &  &  &  &  &  &  &  &  &  &  &  \\
$|22\rangle_1$ &  &  &  & $\frac{3}{4\sqrt{2}}$ & $\frac{\sqrt{3}}{4}$ & $-\frac{1}{4\sqrt{2}}$ &  & $\frac{\sqrt{15}}{8}$
 & $-\frac{1}{8}\sqrt{\frac{3}{5}}$ & $-\frac{9}{8\sqrt{10}}$ & $\frac{1}{8}\sqrt{\frac{3}{10}}$ & $\frac{1}{8\sqrt{5}}$ \\
$|22\rangle_2$ &  &  &  & $\frac{1}{4}\sqrt{\frac{7}{30}}$ & $-\frac{1}{4}\sqrt{\frac{7}{5}}$ & $-\frac{1}{4}\sqrt{\frac{21}{10}}$ & 
 & $-\frac{\sqrt{7}}{40}$ & $-\frac{3\sqrt{7}}{40}$ & $-\frac{1}{40}\sqrt{\frac{21}{2}}$ & $-\frac{1}{8}\sqrt{\frac{7}{2}}$
 & $\frac{\sqrt{21}}{40}$ \\
$|22\rangle_3$ &  &  &  &  & $-\frac{1}{\sqrt{6}}$ & $-\frac{1}{3}$ &  & $-\frac{1}{\sqrt{30}}$ & $-\frac{1}{\sqrt{30}}$ & 
 & $-\frac{1}{\sqrt{15}}$ & $\frac{1}{3\sqrt{10}}$ \\
$|22\rangle_4$ &  &  &  & $\frac{1}{4}$ & $-\frac{1}{2\sqrt{6}}$ & $-\frac{5}{12}$ &  & $\frac{1}{4\sqrt{30}}$
 & $-\frac{1}{4}\sqrt{\frac{5}{6}}$ & $-\frac{3}{8\sqrt{5}}$ & $-\frac{7}{8\sqrt{15}}$ & $\frac{1}{12}\sqrt{\frac{5}{2}}$ \\
$|22\rangle_5$ &  &  &  & $-\frac{1}{2}$ & $-\frac{1}{\sqrt{6}}$ & $\frac{1}{6}$ &  & $-\frac{1}{2}\sqrt{\frac{5}{6}}$ & $\frac{1}{2\sqrt{30}}$
 & $\frac{3}{4\sqrt{5}}$ & $-\frac{1}{4\sqrt{15}}$ & $-\frac{1}{6\sqrt{10}}$ \\
$|22\rangle_6$ &  &  &  & $-\frac{1}{2\sqrt{15}}$ & $\frac{1}{\sqrt{10}}$ & $\frac{1}{2}\sqrt{\frac{3}{5}}$ &  & $\frac{1}{10\sqrt{2}}$
 & $\frac{3}{10\sqrt{2}}$ & $\frac{\sqrt{3}}{20}$ & $\frac{1}{4}$ & $-\frac{1}{10}\sqrt{\frac{3}{2}}$ \\
$|14\rangle_1$ & $\frac{1}{\sqrt{2}}$ & $-\frac{1}{2}\sqrt{\frac{3}{2}}$ & $\frac{1}{2\sqrt{2}}$ & $-\frac{\sqrt{3}}{4}$
 & $\frac{1}{2\sqrt{2}}$ & $-\frac{\sqrt{3}}{4}$ &  & $-\frac{1}{4\sqrt{6}}$ & $-\frac{1}{4}\sqrt{\frac{3}{2}}$
 & $\frac{3}{8}$ & $-\frac{1}{8\sqrt{3}}$ & $\frac{1}{4\sqrt{2}}$ \\
$|14\rangle_2$ & $-\frac{1}{\sqrt{2}}$ & $\frac{1}{2}\sqrt{\frac{3}{2}}$ & $-\frac{1}{2\sqrt{2}}$ & $\frac{\sqrt{3}}{4}$
 & $-\frac{1}{2\sqrt{2}}$ & $\frac{\sqrt{3}}{4}$ &  & $\frac{1}{4\sqrt{6}}$ & $\frac{1}{4}\sqrt{\frac{3}{2}}$
 & $-\frac{3}{8}$ & $\frac{1}{8\sqrt{3}}$ & $-\frac{1}{4\sqrt{2}}$ \\
$|11\rangle_1$ &  &  &  &  &  &  &  & $-\frac{11}{20\sqrt{3}}$ & $-\frac{\sqrt{3}}{20}$ & $-\frac{3}{10\sqrt{2}}$ & $\frac{1}{2\sqrt{6}}$
 & $-\frac{1}{5}$ \\
$|11\rangle_2$ &  &  &  &  &  &  &  & $-\frac{1}{2\sqrt{30}}$ & $-\frac{1}{2\sqrt{30}}$ &  & $\frac{2}{\sqrt{15}}$ & $\frac{1}{\sqrt{10}}$ \\
$|11\rangle_3$ &  &  &  &  &  &  &  & $-\frac{1}{2\sqrt{30}}$ & $-\frac{1}{2}\sqrt{\frac{5}{6}}$ & $-\frac{1}{2\sqrt{5}}$
 & $\frac{1}{2\sqrt{15}}$ &  \\
$|11\rangle_4$ &  &  &  &  &  &  &  & $\frac{1}{\sqrt{6}}$ &  & $\frac{1}{4}$ & $-\frac{1}{4\sqrt{3}}$ & $\frac{1}{2\sqrt{2}}$ \\
$|11\rangle_5$ &  &  &  &  &  &  &  &  & $\frac{1}{\sqrt{30}}$ & $\frac{1}{4\sqrt{5}}$ & $\frac{1}{4}\sqrt{\frac{3}{5}}$
 & $\frac{1}{2\sqrt{10}}$ \\
$|11\rangle_6$ &  &  &  &  &  &  &  &  & $-\frac{1}{\sqrt{10}}$ & $-\frac{1}{4}\sqrt{\frac{3}{5}}$ & $-\frac{3}{4\sqrt{5}}$
 & $-\frac{1}{2}\sqrt{\frac{3}{10}}$ \\
$|11\rangle_7$ &  &  &  &  &  &  &  & $\frac{1}{\sqrt{50}}$ & $-\frac{\sqrt{2}}{5}$ & $\frac{1}{20\sqrt{3}}$ & $\frac{1}{4}$
 & $\frac{3}{10}\sqrt{\frac{3}{2}}$ \\
$|11\rangle_8$ &  &  &  &  &  &  &  & $\frac{1}{4}$ & $\frac{1}{4}$ & $\frac{1}{2\sqrt{6}}$ & $-\frac{1}{2\sqrt{2}}$ &  \\
$|03\rangle_1$ &  & $-\frac{1}{2}\sqrt{\frac{3}{10}}$ & $-\frac{3}{2\sqrt{10}}$ &  &  &  &  & $\frac{1}{4}\sqrt{\frac{3}{5}}$
 & $-\frac{1}{4}\sqrt{\frac{3}{5}}$ & $\frac{3}{4\sqrt{10}}$ & $\frac{1}{4}\sqrt{\frac{3}{10}}$ & $-\frac{3}{4\sqrt{5}}$ \\
$|03\rangle_2$ &  & $\frac{1}{2\sqrt{6}}$ & $\frac{1}{2\sqrt{2}}$ &  &  &  &  & $-\frac{1}{4\sqrt{3}}$ & $\frac{1}{4\sqrt{3}}$
 & $-\frac{1}{4\sqrt{2}}$ & $-\frac{1}{4\sqrt{6}}$ & $\frac{1}{4}$ \\
$|03\rangle_3$ &  & $\frac{1}{2\sqrt{3}}$ & $\frac{1}{2}$ &  &  &  &  & $-\frac{1}{2\sqrt{6}}$ & $\frac{1}{2\sqrt{6}}$ & $-\frac{1}{4}$
 & $-\frac{1}{4\sqrt{3}}$ & $\frac{1}{2\sqrt{2}}$ \\
$|03\rangle_4$ &  & $-\frac{1}{2\sqrt{5}}$ & $-\frac{1}{2}\sqrt{\frac{3}{5}}$ &  &  &  &  & $\frac{1}{2\sqrt{10}}$ & $-\frac{1}{2\sqrt{10}}$
 & $\frac{1}{4}\sqrt{\frac{3}{5}}$ & $\frac{1}{4\sqrt{5}}$ & $-\frac{1}{2}\sqrt{\frac{3}{10}}$ \\
 & & & & & & & & & & & & \Large{(contd.)} \\
\end{tabular}
\end{ruledtabular}

\end{table*}
\begin{table*}
\begin{ruledtabular}
\begin{tabular}{cccccccccccccc}
$Y\,I$ & \multicolumn{2}{c}{$1\,\frac{1}{2}$} & \multicolumn{5}{c}{$1\,\frac{3}{2}$} & $1\,\frac{5}{2}$ & $-1\,\frac{5}{2}$
 & \multicolumn{2}{c}{$-2\,0$} & $-2\,2$ & $-3\,\frac{1}{2}$ \\
\cline{1-1}\cline{2-3}\cline{4-8}\cline{9-9}\cline{10-10}\cline{11-12}\cline{13-13}\cline{14-14}
 & $\Sigma\Lambda N$ & $\Sigma\Lambda N$ & $\Xi NN$ & $\Sigma\Sigma N$ & $\Sigma\Sigma N$
 & $\Sigma\Lambda N$ & $\Sigma\Lambda N$ & $\Sigma\Sigma N$ & $\Xi\Sigma\Sigma$ & $\Xi\Xi\Lambda$ & $\Xi\Xi\Sigma$ & $\Xi\Xi\Sigma$ & $\Xi\Xi\Xi$ \\ 
 & v=1 & v=2 & & v=1 & v=2 & v=1 & v=2 & & & & & & \\
\cline{1-1}\cline{2-3}\cline{4-8}\cline{9-9}\cline{10-10}\cline{11-12}\cline{13-13}\cline{14-14}
$|41\rangle_1$ &  &  & $-\frac{1}{4}\sqrt{\frac{5}{3}}$ & $-\frac{1}{8\sqrt{3}}$ & $-\frac{\sqrt{5}}{8}$
 & $\frac{\sqrt{5}}{8}$ & $-\frac{\sqrt{15}}{8}$ & $\frac{1}{\sqrt{2}}$ &  & $-\frac{1}{2}\sqrt{\frac{3}{2}}$ & $-\frac{1}{2\sqrt{2}}$
 &  & $\frac{1}{\sqrt{2}}$ \\
$|41\rangle_2$ &  &  & $\frac{1}{4}\sqrt{\frac{5}{3}}$ & $\frac{1}{8\sqrt{3}}$ & $\frac{\sqrt{5}}{8}$
 & $-\frac{\sqrt{5}}{8}$ & $\frac{\sqrt{15}}{8}$ & $-\frac{1}{\sqrt{2}}$ &  & $\frac{1}{2}\sqrt{\frac{3}{2}}$ & $\frac{1}{2\sqrt{2}}$
 &  & $-\frac{1}{\sqrt{2}}$ \\
$|30\rangle_1$ &  &  & $-\frac{1}{2}\sqrt{\frac{3}{10}}$ & $\frac{1}{4}\sqrt{\frac{3}{2}}$
 & $\frac{3}{4\sqrt{10}}$ & $-\frac{3}{4\sqrt{10}}$ & $-\frac{1}{4}\sqrt{\frac{3}{10}}$ &  &  & $-\frac{1}{2}\sqrt{\frac{3}{10}}$
 & $\frac{3}{2\sqrt{10}}$ &  &  \\
$|30\rangle_2$ &  &  & $\frac{1}{2\sqrt{6}}$ & $-\frac{1}{4}\sqrt{\frac{5}{6}}$ & $-\frac{1}{4\sqrt{2}}$
 & $\frac{1}{4\sqrt{2}}$ & $\frac{1}{4\sqrt{6}}$ &  &  & $\frac{1}{2\sqrt{6}}$ & $-\frac{1}{2\sqrt{2}}$ &  &  \\
$|30\rangle_3$ &  &  & $-\frac{1}{2\sqrt{3}}$ & $\frac{1}{4}\sqrt{\frac{5}{3}}$ & $\frac{1}{4}$
 & $-\frac{1}{4}$ & $-\frac{1}{4\sqrt{3}}$ &  &  & $-\frac{1}{2\sqrt{3}}$ & $\frac{1}{2}$ &  &  \\
$|30\rangle_4$ &  &  & $-\frac{1}{2\sqrt{5}}$ & $\frac{1}{4}$ & $\frac{1}{4}\sqrt{\frac{3}{5}}$
 & $-\frac{1}{4}\sqrt{\frac{3}{5}}$ & $-\frac{1}{4\sqrt{5}}$ &  &  & $-\frac{1}{2\sqrt{5}}$ & $\frac{1}{2}\sqrt{\frac{3}{5}}$ &  &  \\
$|22\rangle_1$ & $\frac{3}{4\sqrt{5}}$ & $-\frac{1}{8}\sqrt{\frac{3}{5}}$
 & $\frac{1}{4}\sqrt{\frac{3}{2}}$ & $\frac{1}{8}\sqrt{\frac{15}{2}}$ & $-\frac{5}{8\sqrt{2}}$
 & $-\frac{3}{8\sqrt{2}}$ & $-\frac{1}{8}\sqrt{\frac{3}{2}}$ &  &  &  &  &  &  \\
$|22\rangle_2$ & $\frac{1}{20}\sqrt{\frac{7}{3}}$ & $\frac{\sqrt{7}}{8}$
 & $\frac{1}{4}\sqrt{\frac{7}{10}}$ & $\frac{1}{8}\sqrt{\frac{7}{2}}$ & $-\frac{1}{8}\sqrt{\frac{21}{10}}$
 & $\frac{1}{8}\sqrt{\frac{35}{6}}$ & $-\frac{1}{8}\sqrt{\frac{7}{10}}$ &  &  &  &  &  &  \\
$|22\rangle_3$ & & $\sqrt{\frac{2}{15}}$
 & $\frac{1}{4\sqrt{3}}$ & $\frac{1}{8}\sqrt{\frac{5}{3}}$ & $\frac{7}{24}$
 & $\frac{3}{8}$ & $-\frac{1}{8\sqrt{3}}$ &  &  &  &  &  &  \\
$|22\rangle_4$ & $\frac{1}{2\sqrt{10}}$ & $\frac{7}{4\sqrt{30}}$
 & $\frac{1}{2\sqrt{3}}$ & $\frac{1}{4}\sqrt{\frac{5}{3}}$ & $\frac{1}{12}$
 & $\frac{1}{4}$ & $-\frac{1}{4\sqrt{3}}$ &  &  &  &  &  &  \\
$|22\rangle_5$ & $-\frac{1}{\sqrt{10}}$ & $\frac{1}{2\sqrt{30}}$
 & $-\frac{1}{2\sqrt{3}}$ & $-\frac{1}{4}\sqrt{\frac{5}{3}}$ & $\frac{5}{12}$
 & $\frac{1}{4}$ & $\frac{1}{4\sqrt{3}}$ &  &  &  &  &  &  \\
$|22\rangle_6$ & $-\frac{1}{5\sqrt{6}}$ & $-\frac{1}{2\sqrt{2}}$
 & $-\frac{1}{2\sqrt{5}}$ & $-\frac{1}{4}$ & $-\frac{1}{4}\sqrt{\frac{3}{5}}$
 & $-\frac{1}{4}\sqrt{\frac{5}{3}}$ & $\frac{1}{4\sqrt{5}}$ &  &  &  &  &  &  \\
$|14\rangle_1$ & $\frac{1}{2\sqrt{2}}$ & $-\frac{1}{4}\sqrt{\frac{3}{2}}$
 & $-\frac{1}{2\sqrt{3}}$ & $\frac{1}{4}\sqrt{\frac{5}{3}}$ & $-\frac{1}{4}$
 & $\frac{1}{4}$ & $\frac{\sqrt{3}}{4}$ &  & $-\frac{1}{\sqrt{2}}$ &  &  & $\frac{1}{\sqrt{2}}$ &  \\
$|14\rangle_2$ & $-\frac{1}{2\sqrt{2}}$ & $\frac{1}{4}\sqrt{\frac{3}{2}}$
 & $\frac{1}{2\sqrt{3}}$ & $-\frac{1}{4}\sqrt{\frac{5}{3}}$ & $\frac{1}{4}$
 & $-\frac{1}{4}$ & $-\frac{\sqrt{3}}{4}$ &  & $\frac{1}{\sqrt{2}}$ &  &  & $-\frac{1}{\sqrt{2}}$ &  \\
$|11\rangle_1$ & $\frac{1}{5}$ &  &  &  &  &  &  &  &  &  &  &  &  \\
$|11\rangle_2$ &  & $\frac{1}{\sqrt{30}}$ &  &  &  &  &  &  &  &  &  &  &  \\
$|11\rangle_3$ & $-\frac{1}{\sqrt{10}}$ & $-\frac{1}{\sqrt{30}}$ &  &  &  &  &  &  &  &  &  &  &  \\
$|11\rangle_4$ & $-\frac{1}{2\sqrt{2}}$ &  &  &  &  &  &  &  &  &  &  &  &  \\
$|11\rangle_5$ & $\frac{1}{2\sqrt{10}}$ & $\frac{1}{\sqrt{30}}$ &  &  &  &  &  &  &  &  &  &  &  \\
$|11\rangle_6$ & $-\frac{1}{2}\sqrt{\frac{3}{10}}$ & $-\frac{1}{\sqrt{10}}$ &  &  &  &  &  &  &  &  &  &  &  \\
$|11\rangle_7$ & $-\frac{3}{10}\sqrt{\frac{3}{2}}$ &  &  &  &  &  &  &  &  &  &  &  &  \\
$|11\rangle_8$ &  &  &  &  &  &  &  &  &  &  &  &  &  \\
$|03\rangle_1$ &  & $\frac{1}{4}\sqrt{\frac{3}{5}}$ &  &  &  &  &  &  &  &  &  &  &  \\
$|03\rangle_2$ &  & $-\frac{1}{4\sqrt{3}}$ &  &  &  &  &  &  &  &  &  &  &  \\
$|03\rangle_3$ &  & $-\frac{1}{2\sqrt{6}}$ &  &  &  &  &  &  &  &  &  &  &  \\
$|03\rangle_4$ &  & $\frac{1}{2\sqrt{10}}$ &  &  &  &  &  &  &  &  &  &  & 
\end{tabular}
\end{ruledtabular}
\end{table*}

\begin{table*}
\caption{
Same as Table \ref{su3_half} but for $S=3/2$.
}
\label{su3_three-half}
\begin{ruledtabular}
\begin{tabular}{cccccccccccccc}
$Y\,I$ & 2\,0 & 2\,1 & \multicolumn{3}{c}{$1\,\frac{1}{2}$} & \multicolumn{2}{c}{$1\,\frac{3}{2}$} & $0\,2$ & \multicolumn{3}{c}{$-1\,\frac{1}{2}$}
 & $-2\,0$ & $-2\,1$ \\
\cline{1-1}\cline{2-2}\cline{3-3}\cline{4-6}\cline{7-8}\cline{9-9}\cline{10-12}\cline{13-13}\cline{14-14}
 & $\Lambda NN$ & $\Sigma NN$ & $\Xi NN$ & $\Sigma\Sigma N$ & $\Sigma\Lambda N$
 & $\Sigma\Sigma N$ & $\Sigma\Lambda N$ & $\Xi\Sigma N$ & $\Xi\Xi N$ & $\Xi\Sigma\Sigma$ & $\Xi\Sigma\Lambda$
 & $\Xi\Xi\Lambda$ & $\Xi\Xi\Sigma$ \\ 
\cline{1-1}\cline{2-2}\cline{3-3}\cline{4-6}\cline{7-8}\cline{9-9}\cline{10-12}\cline{13-13}\cline{14-14}$|30\rangle_2$ &  &  &  &  &  & $\frac{1}{\sqrt{2}}$
 & $\frac{1}{\sqrt{6}}$ &  & $\frac{1}{\sqrt{3}}$ &  & $\frac{1}{\sqrt{3}}$ & $\sqrt{\frac{2}{3}}$ &  \\
$|30\rangle_3$ &  &  &  &  &  & $\frac{1}{2}$
 & $\frac{1}{2\sqrt{3}}$ &  & $\frac{1}{\sqrt{6}}$ &  & $\frac{1}{\sqrt{6}}$ & $\frac{1}{\sqrt{3}}$ &  \\
$|22\rangle_3$ &  & $\frac{2}{3}$ & $\sqrt{\frac{2}{15}}$ & $\frac{2}{3}\sqrt{\frac{2}{5}}$ & $\sqrt{\frac{2}{15}}$ & $-\frac{1}{3}$
 & $\frac{1}{\sqrt{3}}$ & $-\frac{2}{3}$ & $\sqrt{\frac{2}{15}}$ & $-\frac{2}{3}\sqrt{\frac{2}{5}}$
 & $-\sqrt{\frac{2}{15}}$ &  & $\frac{2}{3}$ \\
$|22\rangle_4$ &  & $-\frac{2}{3}$ & $-\sqrt{\frac{2}{15}}$ & $-\frac{2}{3}\sqrt{\frac{2}{5}}$ & $-\sqrt{\frac{2}{15}}$ & $\frac{1}{3}$
 & $-\frac{1}{\sqrt{3}}$ & $\frac{2}{3}$ & $-\sqrt{\frac{2}{15}}$ & $\frac{2}{3}\sqrt{\frac{2}{5}}$
 & $\sqrt{\frac{2}{15}}$ &  & $-\frac{2}{3}$ \\
$|22\rangle_5$ &  & $-\frac{1}{3}$ & $-\frac{1}{\sqrt{30}}$ & $-\frac{1}{3}\sqrt{\frac{2}{5}}$ & $-\frac{1}{\sqrt{30}}$ & $\frac{1}{6}$
 & $-\frac{1}{2\sqrt{3}}$ & $\frac{1}{3}$ & $-\frac{1}{\sqrt{30}}$ & $\frac{1}{3}\sqrt{\frac{2}{5}}$
 & $\frac{1}{\sqrt{30}}$ &  & $-\frac{1}{3}$ \\
$|11\rangle_2$ &  &  & $\frac{1}{\sqrt{30}}$ & $-\frac{1}{\sqrt{10}}$ & $\frac{1}{\sqrt{30}}$ & 
 &  &  & $\frac{1}{\sqrt{30}}$ & $\frac{1}{\sqrt{10}}$ & $-\frac{1}{\sqrt{30}}$ &  &  \\
$|11\rangle_3$ &  &  & $-\frac{1}{\sqrt{30}}$ & $\frac{1}{\sqrt{10}}$ & $-\frac{1}{\sqrt{30}}$ & 
 &  &  & $-\frac{1}{\sqrt{30}}$ & $-\frac{1}{\sqrt{10}}$ & $\frac{1}{\sqrt{30}}$ &  &  \\
$|11\rangle_5$ &  &  & $-\sqrt{\frac{2}{15}}$ & $\sqrt{\frac{2}{5}}$ & $-\sqrt{\frac{2}{15}}$ & 
 &  &  & $-\sqrt{\frac{2}{15}}$ & $-\sqrt{\frac{2}{5}}$ & $\sqrt{\frac{2}{15}}$ &  &  \\
$|03\rangle_2$ & $\sqrt{\frac{2}{3}}$ &  & $\frac{1}{\sqrt{3}}$ &  & $-\frac{1}{\sqrt{3}}$ &  &  &  &  &  &  &  &  \\
$|03\rangle_3$ & $-\frac{1}{\sqrt{3}}$ &  & $-\frac{1}{\sqrt{6}}$ &  & $\frac{1}{\sqrt{6}}$ &  &  &  &  &  &  &  &  \\
\end{tabular}
\end{ruledtabular}
\end{table*}

\section{\label{sec:level3} QUARK EXCHANGES OF  
THREE OCTET-BARYON STATES}

The 9-quark three-$B_8$ wave function with $SaS_zI_z$ that is antisymmetric 
under the baryon exchange is given by 
\begin{align}
&\Psi_{S aS_zI_z}(B_1B_2B_3,v) \notag \\
&=\Psi^{(\rm orb)}(B_1B_2B_3) \Psi^{(\rm SF)}_{SaS_zI_z}(a_1a_2a_3,v) \Psi^{(\rm color)}(B_1B_2B_3),
\label{3B-basis}
\end{align}
where $\Psi^{(\mbox{\scriptsize{orb}})}(B_1B_2B_3)$ denotes the orbital part with 
the $(0s)^9$ configuration, and $\Psi^{(\mbox{\scriptsize{color}})}(B_1B_2B_3)$ is the color wave function, 
$C(123)C(456)C(789)$. The spin-flavor part $\Psi^{(\rm SF)}_{SaS_zI_z}(a_1a_2a_3,v)$ is 
given by Eq.~(\ref{3B.SF}) with $B_{a_i}$ being replaced by the corresponding 
3-quark wave function $W^{[3]}_{a_i}$. 
In order to examine the quark-Pauli effect,  
we have to solve the eigenvalue problem of the antisymmetrizer ${\cal A}$ 
that makes the 9-quark wave function totally antisymmetric under the exchange of quarks among the baryons. Those eigenfunctions that correspond 
to vanishing eigenvalues or considerably small eigenvalues are 
called Pauli-forbidden or almost Pauli-forbidden. Three $B_8$ baryons can not 
occupy such forbidden configurations, namely they exhibit quark-Pauli repulsion.

It is useful to note that  ${\cal A}$ is scalar with respect to the 
total spin and flavor SU(3) 
label, namely it has no matrix elements between different SU(3) labels  
though it mixes the content of three baryon species. 
This is the reason why we express the basis~(\ref{3B-basis}) in the flavor SU(3) 
representation.

Since we have three baryons each of which consists of three quarks and 
is antisymmetrized, ${\cal A}$ reduces to 55 terms of five basic 
types~\cite{TSH1982}:
\begin{align}
{\cal A}&=\left[1-9(P_{36}+P_{39}+P_{69})+27(P_{369}+P_{396})\right. \nonumber \\
&\ \left.+54(P_{25}P_{39}+P_{35}P_{69}+P_{38}P_{69})\right]
\Big(\sum_{{\cal P}=1}^6(-1)^{\pi(\cal P)}{\cal P}\Big) \nonumber \\
&\ -216P_{25}P_{38}P_{69},
\end{align}
where $P_{ij}$ exchanges quarks $i$ and $j$ and acts on the full orbital, color, spin, and flavor degrees of freedom. The six ${\cal P}$ include those quark exchanges that are equivalent to baryon exchanges. 
Of the five basic types of terms in ${\cal A}$, the first is the direct term, and the second, 
$(P_{36}+P_{39}+P_{69})\left(\sum_{{\cal P}=1}^6(-1)^{\pi(\cal P)}{\cal P}\right)$, involves only 
the exchange of baryons and one quark pair. 
Terms in ${\cal A}$ of the third to fifth category involves the exchange of quark pairs between different baryon pairs. 
We call the first to fifth terms D, 2B, 3Ba, 3Bb, 3Bc in what follows. 

Our three-$B_8$ wave function~(\ref{3B-basis}) is already  
antisymmetrized with respect to the baryon exchanges, so that ${\cal A}$ can be 
effectively replaced by ${\cal A}'=\frac{1}{6}{\cal A}$. 
The matrix elements of ${\cal A}'$ with the basis functions (\ref{3B-basis}) 
are obtained by combining those of the orbital, spin-isospin, and color parts. 
The spin-isospin matrix elements are evaluated by making use of the decomposition~(\ref{3B.SF}) in the flavor SU(3) basis.  
Apparently ${\cal A}'$ conserves $S$ as well as the subgroup labels $I$ and $Y$ of the flavor SU(3) group. 
The orbital matrix element is unity, 
independent of the quark exchange because of the fully symmetric $(0s)^9$ configuration. 
The color matrix element is also simple: the fifth term in ${\cal A}'$, 3Bc, has no color matrix element, 
whereas those of the other terms are $\frac{1}{3},\,\frac{1}{9},\,\frac{1}{9}$, for $P_{36},\,P_{369},\,P_{25}P_{39}$, respectively~\cite{TSH1982}. 
The full matrix elements including the orbital, 
spin-flavor, and color parts between 
the basis functions,
\begin{align}   
\langle S_{12}(\lambda_{12}\mu_{12}){\rho_{12}};S(\lambda\mu)\rho a |  
{\cal A}'  |
S'_{12}(\lambda'_{12}\mu'_{12}){\rho'_{12}};S(\lambda\mu)\rho a\rangle,
\label{ME.A}
\end{align} 
are calculated by the use of SU(3) 6-$(\lambda \mu)$ and 9-$(\lambda \mu)$ 
coefficients~\cite{Millener78} and are tabulated in Appendix C of Supplemental Material~\cite{SM}. 
Assuming the eigenfunction of ${\cal A}'$ to be  
\begin{align}
\sum_{B_1B_2B_3 v}C_{Sa}(B_1B_2B_3,v) \Psi_{SaS_zI_z}(B_1B_2B_3,v), 
\end{align}
we solve the eigenvalue problem 
\begin{widetext}
\begin{eqnarray}
\sum_{B_1^\prime B_2^\prime B_3^\prime v'}
\langle \Psi_{SaS_zI_z}(B_1B_2B_3,v)\big|
{\cal A}'
\big|\Psi_{SaS_zI_z}(B_1^\prime B_2^\prime B_3^\prime,v^\prime)\rangle
C_{Sa}(B_1^\prime B_2^\prime B_3^\prime, v')
=\mu_{Sa}C_{Sa}(B_1B_2B_3, v).
\label{norm-eigenvalue-eq}
\end{eqnarray}
\end{widetext}
The eigenvalues $\mu_{Sa}$ are given in Tables \ref{eigenvalue_half} and \ref{eigenvalue_three-half}. The eigenfunctions are available from the authors of this paper 
upon request.

\begin{table*}
\caption{Eigenvalues $\mu_{Sa}$ in Eq.~(\ref{norm-eigenvalue-eq}), given in 
increasing order, for 
three-$B_8$ systems with $S=1/2$. The expectation value of ${\cal A}'$ calculated 
for each $B_8B_8B_8$ system is given in $\langle{\cal A}'\rangle$ column.  }
\label{eigenvalue_half}
\begin{ruledtabular}
{\renewcommand\arraystretch{1.7}
\begin{tabular}{clcccclcccclcc}
$Y\,I$ & $B_8B_8B_8$ & $\langle{\cal A}'\rangle$ & $\mu_{Sa}$
 && $Y\,I$ & $B_8B_8B_8$ & $\langle{\cal A}'\rangle$ &  $\mu_{Sa}$ 
 && $Y\,I$ & $B_8B_8B_8$ & $\langle{\cal A}'\rangle$ &  $\mu_{Sa}$ \\
\cline{1-4}\cline{6-9}\cline{11-14}
$3\,\frac{1}{2}$ & $NNN$ & $\frac{100}{81}$ & $\frac{100}{81}$
 && 0\,0 & $\Xi\Lambda N \,{v=1}$ & $\frac{5}{6}$ & 0
 && $-1\,\frac{1}{2}$ & $\Xi\Xi N \,{v=1}$ & $\frac{17}{81}$ & 0 \\
\cline{1-4}
2\,0 & $\Lambda NN$ & $\frac{25}{27}$ & 0 
 &&  & $\Xi\Lambda N \,{v=2}$ & $\frac{55}{54}$ & 0
 &&  & $\Xi\Xi N \,{v=2}$ & $\frac{73}{81}$ & 0 \\
 & $\Sigma NN$ & $\frac{25}{81}$ &  $\frac{100}{81}$
 &&  & $\Xi\Sigma N \,{v=1}$ & $\frac{5}{54}$ & 0
 &&  & $\Xi\Lambda\Lambda$ & $\frac{1}{2}$ & 0 \\
\cline{1-4}
2\,1 & $\Lambda NN$ & $\frac{25}{27}$ & 0
 &&  & $\Xi\Sigma N \,{v=2}$ & $\frac{55}{486}$ & $\frac{200}{243}$
 &&  & $\Xi\Sigma\Sigma \,{v=1}$ & $\frac{83}{162}$ & 0 \\
 & $\Sigma NN\, {v=1}$ & $\frac{50}{81}$ & $\frac{200}{243}$
 &&  & $\Sigma\Sigma\Lambda$ & $\frac{10}{27}$ & $\frac{130}{81}$
 &&  & $\Xi\Sigma\Sigma \,{v=2}$ & $\frac{17}{243}$ & $\frac{4}{81}$ \\
\cline{6-9}
& $\Sigma NN \,{v=2}$ & $\frac{125}{243}$ & $\frac{100}{81}$
 && 0\,1 & $\Xi\Lambda N \,{v=1}$ & $\frac{33}{54}$ & 0
 &&  & $\Xi\Sigma\Lambda \,{v=1}$ & $\frac{1}{36}$ & $\frac{200}{243}$ \\
\cline{1-4}
2\,2 & $\Sigma NN$ & $\frac{4}{81}$ & $\frac{4}{81}$
 &&  & $\Xi\Lambda N \,{v=2}$ & $\frac{17}{162}$ & 0
 &&  & $\Xi\Sigma\Lambda \,{v=2}$ & $\frac{83}{324}$ & $\frac{130}{81}$ \\
\cline{1-4}\cline{11-14}
$1\,\frac{1}{2}$ & $\Xi NN\, {v=1}$ & $\frac{25}{27}$ & 0
 &&  & $\Xi\Sigma N \,{v=1}$ & $\frac{11}{162}$ & 0
 && $-1\,\frac{3}{2}$ & $\Xi\Xi N$ & $\frac{34}{81}$ & 0 \\
 & $\Xi NN\, {v=2}$ & $\frac{35}{81}$ & 0
 &&  & $\Xi\Sigma N \,{v=2}$ & $\frac{17}{27}$ & 0
 &&  & $\Xi\Sigma\Sigma\,{v=1}$ & $\frac{35}{162}$ & 0 \\
 & $\Lambda\Lambda N$ & $\frac{5}{6}$ & 0
 &&  & $\Xi\Sigma N \,{v=3}$ & $\frac{673}{1458}$ & 0
 &&  & $\Xi\Sigma\Sigma\,{v=2}$ & $\frac{253}{486}$ & $\frac{4}{81}$ \\
 & $\Sigma\Sigma N \,{v=1}$ & $\frac{85}{162}$ & 0
 &&  & $\Xi\Sigma N \,{v=4}$ & $\frac{295}{729}$ & $\frac{4}{81}$
 &&  & $\Xi\Sigma\Lambda\,{v=1}$ & $\frac{1}{3}$ & $\frac{200}{243}$ \\
 & $\Sigma\Sigma N \,{v=2}$ & $\frac{35}{243}$ & $\frac{200}{243}$
 &&  & $\Sigma\Lambda\Lambda$ & $\frac{4}{9}$ & $\frac{200}{243}$
 &&  & $\Xi\Sigma\Lambda\,{v=2}$ & $\frac{50}{81}$ & $\frac{100}{81}$ \\
\cline{11-14}
 & $\Sigma\Lambda N \,{v=1}$ & $\frac{5}{9}$ & $\frac{100}{81}$
 &&  & $\Sigma\Sigma\Lambda$ & $\frac{23}{81}$ & $\frac{100}{81}$
 && $-1\,\frac{5}{2}$ & $\Xi\Sigma\Sigma$ & $\frac{100}{81}$ & $\frac{100}{81}$ \\
\cline{11-14}
 & $\Sigma\Lambda N\, {v=2}$ & $\frac{20}{81}$ & $\frac{130}{81}$
 &&  & $\Sigma\Sigma\Sigma$ & $\frac{19}{27}$ & $\frac{130}{81}$
 && $-2$\,0 & $\Xi\Xi\Lambda$ & $\frac{1}{27}$ & 0 \\
\cline{1-4}\cline{6-9}
$1\,\frac{3}{2}$ & $\Xi NN$ & $\frac{10}{27}$ & 0
 && 0\,2 & $\Xi\Sigma N\,{v=1}$ & $\frac{1}{3}$ & 0
 &&  & $\Xi\Xi\Sigma$ & $\frac{1}{81}$ & $\frac{4}{81}$ \\
\cline{11-14}
 & $\Sigma\Sigma N\, {v=1}$ & $\frac{73}{162}$ & 0
 &&  & $\Xi\Sigma N\,{v=2}$ & $\frac{125}{243}$ & $\frac{4}{81}$
 && $-2\,1$ & $\Xi\Xi\Lambda$ & $\frac{13}{27}$ & 0 \\
 & $\Sigma\Sigma N\, {v=2}$ & $\frac{235}{486}$ & $\frac{4}{81}$
 &&  & $\Sigma\Sigma\Lambda$ & $\frac{13}{27}$ & $\frac{200}{243}$
 &&  & $\Xi\Xi\Sigma\,{v=1}$ & $\frac{26}{81}$ & $\frac{4}{81}$ \\
 & $\Sigma\Lambda N\, {v=1}$ & $\frac{5}{18}$ & $\frac{200}{243}$
 &&  & $\Sigma\Sigma\Sigma$ & $\frac{7}{9}$ & $\frac{100}{81}$
 &&  & $\Xi\Xi\Sigma\,{v=2}$ & $\frac{17}{243}$ & $\frac{200}{243}$ \\
\cline{11-14}
 & $\Sigma\Lambda N\, {v=2}$ & $\frac{85}{162}$ & $\frac{100}{81}$
 &&  &  &  & 
 && $-2\,2$ & $\Xi\Xi\Sigma$ & $\frac{100}{81}$ & $\frac{100}{81}$ \\
\cline{1-4}\cline{11-14}
$1\,\frac{5}{2}$ & $\Sigma\Sigma N$ & $\frac{4}{81}$ & $\frac{4}{81}$
 &&  &  &  & 
 && $-3\,\frac{1}{2}$ & $\Xi\Xi\Xi$ & $\frac{4}{81}$ & $\frac{4}{81}$ \\
\end{tabular}}
\end{ruledtabular}
\end{table*}
\begin{table}
\caption{Same as Table \ref{eigenvalue_half} but for $S=3/2$.
}
\label{eigenvalue_three-half}
\begin{ruledtabular}
{\renewcommand\arraystretch{1.7}
\begin{tabular}{cclcc}
$Y$ & $I$ & $B_8B_8B_8$ & $\langle{\cal A}'\rangle$ & $\mu_{Sa}$ \\
\hline
2 & 0 & $\Lambda NN$ & $\frac{25}{27}$ & $\frac{25}{27}$  \\
\hline
2 & 1 & $\Sigma NN$ & $\frac{35}{243}$ &  $\frac{35}{243}$ \\
\hline
1 & $\frac{1}{2}$ & $\Xi NN$ & $\frac{50}{81}$ & $\frac{35}{243}$ \\
 & & $\Sigma\Sigma N$ & $\frac{95}{243}$ & $\frac{5}{9}$ \\
 & & $\Sigma\Lambda N$ & $\frac{50}{81}$ & $\frac{25}{27}$ \\
\hline
1 & $\frac{3}{2}$ & $\Sigma\Sigma N$ & $\frac{31}{486}$ & $\frac{1}{27}$  \\
 & & $\Sigma\Lambda N$ & $\frac{19}{162}$ & $\frac{35}{243}$ \\
\hline
0 & 0 & $\Xi\Lambda N$ & $\frac{20}{27}$ & $\frac{35}{243}$ \\
& & $\Xi\Sigma N$ & $\frac{140}{243}$ & $\frac{5}{9}$ \\
& & $\Sigma\Sigma\Sigma$ & $\frac{55}{81}$ & $\frac{35}{27}$ \\
\hline
0 & 1 & $\Xi\Lambda N$ & $\frac{34}{81}$ & $\frac{1}{27}$ \\
 & & $\Xi\Sigma N\, {v=1}$ & $\frac{134}{729}$ & $\frac{35}{243}$ \\
 & & $\Xi\Sigma N\, {v=2}$ & $\frac{565}{729}$ & $\frac{5}{9}$ \\
 & & $\Sigma\Sigma\Lambda$ & $\frac{23}{81}$ & $\frac{25}{27}$ \\
\hline
0 & 2 & $\Xi\Sigma N$ & $\frac{35}{243}$ & $\frac{35}{243}$ \\
\hline
$-1$ & $\frac{1}{2}$ & $\Xi\Xi N$ & $\frac{14}{81}$ & $\frac{1}{27}$ \\
 & & $\Xi\Sigma\Sigma$ & $\frac{95}{243}$ & $\frac{35}{243}$ \\
 & & $\Xi\Sigma\Lambda$ & $\frac{14}{81}$ & $\frac{5}{9}$ \\
\hline
$-1$ & $\frac{3}{2}$ & $\Xi\Sigma\Sigma$ & $\frac{355}{486}$ & $\frac{35}{243}$ \\
 & & $\Xi\Sigma\Lambda$ & $\frac{55}{162}$ &  $\frac{25}{27}$ \\
\hline
$-2$ & 0 & $\Xi\Xi\Lambda$ & $\frac{1}{27}$ & $\frac{1}{27}$ \\
\hline
$-2$ & 1 & $\Xi\Xi\Sigma$ & $\frac{35}{243}$ & $\frac{35}{243}$ \\
\end{tabular}}
\end{ruledtabular}
\end{table}

As three-$B_8$ systems that do not couple with other systems, 
we have six and five systems in the $S=\frac{1}{2}$ and $\frac{3}{2}$ cases, 
respectively. They are 
$NNN(I=\frac{1}{2})$, 
$\Sigma NN(2)$, $\Sigma \Sigma N(\frac{5}{2})$, 
$\Xi\Sigma\Sigma(\frac{5}{2})$, $\Xi\Xi \Sigma(2)$, 
$\Xi \Xi \Xi(\frac{1}{2})$ for $S=\frac{1}{2}$ and $\Lambda NN(0)$, $\Sigma NN(1)$, 
$\Xi \Sigma N(2)$, $\Xi \Xi \Lambda(0)$, $\Xi \Xi \Sigma (1)$ for $S=\frac{3}{2}$, respectively. Among these, $\Sigma NN(2)$, $\Sigma \Sigma N(\frac{5}{2})$, 
$\Xi \Xi \Xi(\frac{1}{2})$, and $\Sigma NN(1)$, $\Xi \Sigma N(2)$, 
$\Xi \Xi \Lambda(0)$, $\Xi \Xi \Sigma (1)$ are all considered  
almost Pauli-forbidden states because the corresponding $\mu_{Sa}$ values are 
fairly small. 
The strong quark-Pauli repulsion in the $S=\frac{1}{2}$ systems,   
$\Sigma NN(2)$, $\Sigma\Sigma N(\frac{5}{2})$, and $\Xi\Xi\Xi(\frac{1}{2})$, 
results from the fact 
that they consist of the $|41\rangle_1$ and $|41\rangle_2$ components 
(see Table~\ref{su3_half}) whose matrix elements of ${\cal A}'$ 
\begin{align}
\left(
\begin{array}{cc}
\frac{2}{81} & -\frac{2}{81} \\
-\frac{2}{81} & \frac{2}{81} \\
\end{array}
\right)
\end{align}
are small. The reason that the above-mentioned systems with $S=\frac{3}{2}$ become 
almost Pauli-forbidden is similar to the $S=\frac{1}{2}$ case. 
As seen in Table~\ref{su3_three-half}, $\Sigma NN(1)$, $\Xi\Sigma N(2)$ and $\Xi\Xi\Sigma(1)$ consist of $|22\rangle_n$-components 
and $\Xi\Xi\Lambda(0)$ consists of $|30\rangle_n$-components. 
The matrix elements of those components are relatively small as given in  
Appendix C of Supplemental Material~\cite{SM}.

In the coupled-channel three-$B_8$ case, 
we find many completely Pauli-forbidden states $\Phi^{\rm FS}_{Sa}$. 
Examples of $\Phi^{\rm FS}_{Sa}$ that are uniquely determined are 
\begin{align}
&\Phi^{\rm FS}_{ \frac{1}{2}\,2\,0}={\textstyle{\frac{1}{2}}}\Psi_{{\scriptsize \frac{1}{2}}\,2\,0}(\Lambda NN)
+{\textstyle{\frac{\sqrt{3}}{2}}}\Psi_{{\scriptsize \frac{1}{2}}\,2\,0}(\Sigma NN), 
\notag \\
&\Phi^{\rm FS}_{\frac{1}{2}\,2\,1}=-{\textstyle{\frac{1}{4}}}\Psi_{{\scriptsize \frac{1}{2}}\,2\,1}(\Lambda NN) \notag \\
&\quad +{\textstyle{\sqrt{\frac{3}{8}}}}\Psi_{{\scriptsize \frac{1}{2}}\,2\,1}(\Sigma NN,v\!=\!1)
+{\textstyle{\frac{3}{4}}}\Psi_{{\scriptsize \frac{1}{2}}\,2\,1}(\Sigma NN,v\!=\!2), 
\notag \\
&\Phi^{\rm FS}_{\frac{1}{2}\,0\,2}={\textstyle{\frac{\sqrt{3}}{4}}}\Psi_{{\scriptsize \frac{1}{2}}\,0\,2}(\Xi\Sigma N,v\!=\!1) +{\textstyle{\frac{3}{4}}}\Psi_{{\scriptsize \frac{1}{2}}\,0\,2}(\Xi\Sigma N,v\!=\!2) \notag \\
&\quad  -{\textstyle{\frac{1}{2}}}\Psi_{{\scriptsize \frac{1}{2}}\,0\,2}(\Sigma\Sigma\Lambda), 
\notag \\
&\Phi^{\rm FS}_{\frac{1}{2}\, -\!2\,0}={\textstyle{\frac{1}{2}}}\Psi_{{\scriptsize \frac{1}{2}}\,-2\,0}(\Xi\Xi\Lambda)
-{\textstyle{\frac{\sqrt{3}}{2}}}\Psi_{{\scriptsize \frac{1}{2}}\,-2\,0}(\Xi\Xi\Sigma), 
\notag \\
&\Phi^{\rm FS}_{\frac{1}{2}\, -\!2\,1}={\textstyle{\frac{1}{4}}}\Psi_{{\scriptsize \frac{1}{2}}\,-2\,1}(\Xi\Xi\Lambda) \notag \\
&\quad +{\textstyle{\sqrt{\frac{3}{8}}}}\Psi_{{\scriptsize \frac{1}{2}}\,-2\,1}(\Xi\Xi\Sigma,v\!=\!1)
+{\textstyle{\frac{3}{4}}}\Psi_{{\scriptsize \frac{1}{2}}\,-2\,1}(\Xi\Xi\Sigma,v\!=\!2). 
\label{CPFS.CC}
\end{align}
The existence of these Pauli-forbidden states indicates that they are 
not allowed to take the $(0s)^9$ configuration in, e.g., few-body 
calculations consisting of $N$'s and $Y$'s. 
Other completely Pauli-forbidden states have degeneracy. 
In the coupled-channel three-$B_8$ systems, we also find several 
almost Pauli-forbidden states, for example,  $\Xi \Xi \Lambda +  
\Xi \Xi \Sigma$ with $SI=\frac{1}{2}0$ and 
$\Sigma\Lambda N + \Sigma\Sigma N$ with $SI=\frac{3}{2}\frac{3}{2}$.

Whether or not the $\Lambda NN$ system receives a strong quark-Pauli 
repulsion is particularly interesting for the neutron star problem as noted in 
Introduction. The $\Lambda NN$ system takes $I=0, 1$ in  
$S=\frac{1}{2}$ and  $I=0$ in $S=\frac{3}{2}$. In the latter case the value of 
$\mu_{Sa}$ is large, indicating that the $\Lambda NN$ system with 
$SI=\frac{3}{2}0$ is completely 
allowed. In the former case we have one completely Pauli-forbidden state in 
both  $I=0, 1$ states. The probability of finding the $\Lambda NN$ 
state in those Pauli-forbidden states is, however, rather small as shown 
in Eq.~(\ref{CPFS.CC}), so that the $\Lambda NN$ system receives a 
minor quark-Pauli repulsion. Thus the quark-Pauli repulsion is unlikely 
to suppress the emergence of the $\Lambda$ hyperon that is a candidate for
the first hyperon in increasing baryon density.

In contrast to $\Lambda$, 
the $\Sigma$ hyperon is involved in producing several almost 
Pauli-forbidden $\Sigma NN$ states. A specific case is $\Sigma NN(I=2)$ 
including the $\Sigma^-nn$ state. 
The quark-Pauli effect prevents $\Sigma^-$ from appearing 
with increasing baryon density in the neutron star.

The $H$-dibaryon is conjectured to be a 
$B_8B_8$ bound or resonant state with $S=0, Y=-1, I=0$. 
Its flavor symmetry is 
SU(3)-scalar, that is, $(\lambda_{12}\mu_{12})=(00)$ in 
Eq.~(\ref{SU3_coupling}). No clear experimental confirmation of the 
$H$-dibaryon has been made yet. A question arises if the quark-Pauli 
effect demolishes the most compact three-$B_8$ configuration consisting 
of $HB_8$ if the $H$ is assumed to have a main configuration of $(0s)^6$. 
The $HB_8$ state should have $(\lambda \mu)=(11)$ and 
$S=\frac{1}{2}$. We identify the $HB_8$ state as $|11\rangle_8$ 
in Table~\ref{short_sf}. 
The expectation value of ${\cal A}'$ with this state is 
$\frac{5}{12}$ as seen in the table of Appendix C in 
Supplemental Material~\cite{SM}. 
This implies somewhat repulsive quark-Pauli effect, 
which is qualitatively consistent with the corresponding result of Ref.~\cite{SAKAI2000}.

\section{conclusion}
\label{sec:level4}
 
The hyperons appear to be present in the interior of the neutron star 
with the increasing baryon density. Since their appearance generally 
leads to the softening of its equation of state, some repulsive 
mechanism to suppress the role of the hyperons is called for in order to 
be consistent with the observation that the mass of the neutron star can be 
twice as heavy as the solar mass. At present the information on 
two- and three-baryon forces including the hyperons is very much limited, 
and it is hard to draw clear conclusions on the required repulsion. 
In this work we have considered the role of 
the quark-Pauli blocking effect on the 
three octet-baryons. 
 
Assuming a common orbital wave function for 
all the octet-baryons, we have considered the most compact 
spatial configuration in which the three particles are located on top of 
each other, as possible quark-Pauli effect becomes largest. 
We have first constructed all possible states with the total 
spin $S=\frac{1}{2}$ and 
$\frac{3}{2}$ that are antisymmetric in the simultaneous exchange of 
spin and flavor degrees of freedom  
of the baryons. The flavor SU(3) symmetry of the three particles is 
exploited to classify the constructed states. The quark-Pauli 
effect is then quantified by calculating the eigenvalues of 
the nine-quark antisymmetrizer in those three-baryon states. 

Several systems have been found to have vanishing or small eigenvalues  
that lead to the strong quark-Pauli repulsion.   
In the $S=\frac{1}{2}$ case, they are $\Sigma NN(I=2)$, 
$\Sigma\Sigma N(\frac{5}{2})$, $\Xi\Xi\Xi(\frac{1}{2})$,  
$\Xi\Xi\Lambda(0)$, and $\Xi\Xi\Sigma(0)$, where $I$ is the 
total isospin of the 
three baryons. In the $S=\frac{3}{2}$ case, they are  
$\Sigma NN(1)$, $\Xi\Sigma N(2)$, $\Xi\Xi\Lambda(0)$, $\Xi\Xi\Sigma(1)$, 
$\Sigma\Lambda N(\frac{3}{2})$, and $\Sigma\Sigma N(\frac{3}{2})$. 
The $\Lambda$ and $\Sigma$ hyperons behave differently with 
respect to the quark-Pauli 
repulsion. The $\Lambda NN$ system receives minor quark-Pauli effects and 
are allowed to be present in the interior of the neutron star unless 
$\Lambda NN$ three-body force is strongly repulsive, whereas 
the $\Sigma NN(I=2)$ system regardless of $S=\frac{1}{2}$ or $\frac{3}{2}$, 
including, e.g., $\Sigma^- nn$ is almost Pauli-forbidden.  

It will be interesting to study three octet-baryon forces using  
a quark-model Hamiltonian.  
The spin and flavor SU(3) symmetry developed here should be useful 
to the extent to which the underlying Hamiltonian is  
SU(3)-scalar.  Work along this direction is in progress.

\begin{acknowledgments}
This work was supported in part by JSPS KAKENHI Grants, 24654071, 24540261, and 15K05072.
\end{acknowledgments}


\bibliography{apssamp}

\end{document}


\preprint{APS/123-QED}

\title{Quark-Pauli effects in the three-baryons\\
Supplemental Material}

\author{C. Nakamoto$^1$}
\author{Y. Suzuki$^{2,\,3}$}%
\affiliation{%
$^1$National Institute of Technology, Suzuka College, Suzuka 510-0294, Japan\\
$^2$Department of Physics, Niigata University, Niigata 950-2181, Japan\\
$^3$RIKEN Nishina Center, Wako 351-0198, Japan
}%




\date{\today}


\maketitle


\begin{widetext}


\appendix


\section{Antisymmetric three octet-baryon spin-flavor functions}

\begin{table*}[h]
\caption{Antisymmetric three octet-baryon spin-flavor functions with $S=1/2$ in the spin-isospin basis. 
The three-baryon state $[B_{a_1}B_{a_2}]_{S'a'}B_{a_3}$ stands for 
$[[B_{a_1}(1)B_{a_2}(2)]_{S'a'}B_{a_3}(3)]_{Sa}$. 
Note, however, that the hypercharge is abbreviated for the sake of simplicity.
The label $v$ distinguishes the multiple occurence of the 
orthogonal antisymmetric states for a given $SYI$. This table continues.}
\begin{ruledtabular}
\begin{tabular}{cccclcclc}
$Y$ & $I$ &&& $B_8B_8B_8$ &&&  \\
\hline
3 & $\frac{1}{2}$ &&& $NNN$ &&& $\frac{1}{\sqrt{2}}[NN]_{01}N-\frac{1}{\sqrt{2}}[NN]_{10}N$ \\
\hline
2 &  0    &&& $\Lambda NN$ &&& $\frac{1}{\sqrt{3}}[NN]_{10}\Lambda
          -\frac{1}{\sqrt{2}}[\Lambda N]_{0  \frac{1}{2}}N-\frac{1}{\sqrt{6}}[\Lambda N]_{1\frac{1}{2}}N$ \\ 
 &      &&& $\Sigma NN$ &&& $\frac{1}{\sqrt{3}}[NN]_{01}\Sigma -\frac{1}{\sqrt{6}}
             [\Sigma N]_{0 \frac{1}{2}}N +\frac{1}{\sqrt{2}}[\Sigma N]_{1\frac{1}{2}}N$ \\ 
\hline
2 & 1 &&& $\Lambda NN$ &&& $\frac{1}{\sqrt{3}}[NN]_{01}\Lambda
       -\frac{1}{\sqrt{6}}[\Lambda N]_{0   \frac{1}{2}}N +\frac{1}{\sqrt{2}}[\Lambda N]_{1\frac{1}{2}}N$ \\ 
  &   &&& $\Sigma NN\,{v=1}$ &&& $\frac{1}{\sqrt{3}}[NN]_{01}\Sigma +\frac{1}{3}
            [\Sigma N]_{0 \frac{1}{2}}N +\frac{1}{\sqrt{18}}[\Sigma N]_{0\frac{3}{2}}N -\frac{1}{\sqrt{3}}
            [\Sigma N]_{1\frac{1}{2}}N -\frac{1}{\sqrt{6}}[\Sigma N]_{1\frac{3}{2}}N$ \\ 
  &   &&& $\Sigma NN\,{v=2}$  &&& $\frac{1}{\sqrt{3}}[NN]_{10}\Sigma +\frac{1}{\sqrt{6}}[\Sigma N]_{0 \frac{1}{2}}N -\frac{1}{\sqrt{3}}
            [\Sigma N]_{0\frac{3}{2}}N +\frac{1}{\sqrt{18}}[\Sigma N]_{1\frac{1}{2}}N -\frac{1}{3}[\Sigma N]_{1\frac{3}{2}}N$ \\  
\hline
2 & 2 &&& $\Sigma NN$ &&& $\frac{1}{\sqrt{3}}[NN]_{01}\Sigma 
          -\frac{1}{\sqrt{6}}[\Sigma N]_{0  \frac{3}{2}}N +\frac{1}{\sqrt{2}}[\Sigma N]_{1\frac{3}{2}}N$ \\ 
\hline
1  & $\frac{1}{2}$  &&& $\Xi NN\,{v=1}$ &&& $\frac{1}{\sqrt{3}}[NN]_{01}\Xi +\frac{1}{\sqrt{8}}[\Xi N]_{00}N +\frac{1}{\sqrt{24}} 
                  [\Xi N]_{01}N -\sqrt{\frac{3}{8}}[\Xi N]_{10}N -\frac{1}{\sqrt{8}}[\Xi N]_{11}N $ \\
              &   &&& $\Xi NN\,{v=2}$ &&& $\frac{1}{\sqrt{3}}[NN]_{10}\Xi +\frac{1}{\sqrt{8}}[\Xi N]_{00}N -\sqrt{\frac{3}{8}} 
                  [\Xi N]_{01}N +\frac{1}{\sqrt{24}}[\Xi N]_{10}N -\frac{1}{\sqrt{8}}[\Xi N]_{11}N $ \\ 
              &  &&& $\Lambda \Lambda N$ &&& $\frac{1}{\sqrt{3}}
                 [\Lambda \Lambda ]_{00}\Lambda -\frac{1}{\sqrt{6}}[\Lambda N]_{0\frac{1}{2}}\Lambda -\frac{1}{\sqrt{2}}
                 [\Lambda N]_{1\frac{1}{2}}\Lambda $ \\
              &   &&& $\Sigma \Sigma N\,{v=1}$ &&& $\frac{1}{\sqrt{3}}[\Sigma \Sigma]_{00}N -\frac{1}{\sqrt{18}}
                  [\Sigma N]_{0\frac{1}{2}}\Sigma -\frac{1}{3}[\Sigma N]_{0\frac{3}{2}}\Sigma -\frac{1}{\sqrt{6}} 
                  [\Sigma N]_{1\frac{1}{2}}\Sigma -\frac{1}{\sqrt{3}} [\Sigma N]_{1\frac{3}{2}}\Sigma $ \\
              &   &&& $\Sigma \Sigma N\,{v=2}$ &&& $\frac{1}{\sqrt{3}}[\Sigma \Sigma]_{11}N -\frac{1}{\sqrt{3}}
               [\Sigma N]_{0\frac{1}{2}}\Sigma +\frac{1}{\sqrt{6}} [\Sigma N]_{0\frac{3}{2}}\Sigma +\frac{1}{3} 
                 [\Sigma N]_{1\frac{1}{2}}\Sigma -\frac{1}{\sqrt{18}} [\Sigma N]_{1\frac{3}{2}}\Sigma $ \\ 
   &  &&& $\Sigma \Lambda N\,{v=1}$ &&& $\frac{1}{\sqrt{3}}[\Sigma \Lambda]_{01}N +\frac{1}{\sqrt{12}}
               [\Lambda N]_{0\frac{1}{2}}\Sigma  +\frac{1}{2} [\Lambda N]_{1\frac{1}{2}}\Sigma 
                -\frac{1}{\sqrt{12}} [\Sigma N]_{0\frac{1}{2}}\Lambda -\frac{1}{2} [\Sigma N]_{1\frac{1}{2}}\Lambda $ \\ 
   &  &&& $\Sigma \Lambda N\,{v=2}$ &&& $\frac{1}{\sqrt{3}}[\Sigma \Lambda]_{11}N -\frac{1}{2} [\Lambda N]_{0\frac{1}{2}}\Sigma  +\frac{1}{\sqrt{12}} 
                [\Lambda N]_{1\frac{1}{2}}\Sigma -\frac{1}{2} [\Sigma N]_{0\frac{1}{2}}\Lambda 
                 +\frac{1}{\sqrt{12}} [\Sigma N]_{1\frac{1}{2}}\Lambda $ \\                    
\hline
1  & $\frac{3}{2}$  &&&  $\Xi NN$ &&&  $\frac{1}{\sqrt{3}}[NN]_{01}\Xi -\frac{1}{\sqrt{6}}[\Xi N]_{01}N +\frac{1}{\sqrt{2}} 
                      [\Xi N]_{11}N $ \\
              &   &&& $\Sigma \Sigma N\,{v=1}$ &&& $\frac{1}{\sqrt{3}}[\Sigma \Sigma]_{02}N -\frac{\sqrt{5}}{6}
                   [\Sigma N]_{0\frac{1}{2}}\Sigma +\frac{1}{6} [\Sigma N]_{0\frac{3}{2}}\Sigma -\sqrt{\frac{5}{12}} 
                   [\Sigma N]_{1\frac{1}{2}}\Sigma +\frac{1}{\sqrt{12}} [\Sigma N]_{1\frac{3}{2}}\Sigma $ \\
              &  &&& $\Sigma \Sigma N\,{v=2}$ &&& $\frac{1}{\sqrt{3}}[\Sigma \Sigma]_{11}N -\frac{1}{\sqrt{12}}
                  [\Sigma N]_{0\frac{1}{2}}\Sigma -\sqrt{\frac{5}{12}} [\Sigma N]_{0\frac{3}{2}}\Sigma +\frac{1}{6} 
                  [\Sigma N]_{1\frac{1}{2}}\Sigma +\frac{\sqrt{5}}{6} [\Sigma N]_{1\frac{3}{2}}\Sigma $ \\
             &  &&& $\Sigma \Lambda N\,{v=1}$ &&& $\frac{1}{\sqrt{3}}[\Sigma \Lambda]_{01}N -\frac{1}{\sqrt{12}}
               [\Lambda N]_{0\frac{1}{2}}\Sigma  -\frac{1}{2}  [\Lambda N]_{1\frac{1}{2}}\Sigma 
                 -\frac{1}{\sqrt{12}} [\Sigma N]_{0\frac{3}{2}}\Lambda -\frac{1}{2} [\Sigma N]_{1\frac{3}{2}}\Lambda $ \\  
            &  &&& $\Sigma \Lambda N\,{v=2}$ &&& $\frac{1}{\sqrt{3}}[\Sigma \Lambda]_{11}N +\frac{1}{2} [\Lambda N]_{0\frac{1}{2}}\Sigma  -\frac{1}{\sqrt{12}} 
                [\Lambda N]_{1\frac{1}{2}}\Sigma -\frac{1}{2} [\Sigma N]_{0\frac{3}{2}}\Lambda +\frac{1}{\sqrt{12}} 
                 [\Sigma N]_{1\frac{3}{2}}\Lambda $ \\       
\hline
1 & $\frac{5}{2}$  &&& $\Sigma \Sigma N$ &&& $\frac{1}{\sqrt{3}}[\Sigma \Sigma]_{02}N -\frac{1}{\sqrt{6}}
             [\Sigma N]_{0\frac{3}{2}}\Sigma -\frac{1}{\sqrt{2}} [\Sigma N]_{1\frac{3}{2}}\Sigma $ \\
 & & & & & & & \hspace{80mm} {\Large (contd.)}
\end{tabular}
\end{ruledtabular}
\end{table*}
\begin{table*}
\begin{ruledtabular}
\begin{tabular}{cccclcclc}
 0 & 0 &&& $\Xi \Lambda N\,{v=1}$ &&& $\frac{1}{\sqrt{3}}[\Xi \Lambda]_{0\frac{1}{2}}N +\frac{1}{\sqrt{12}}
                   [\Lambda N]_{0\frac{1}{2}}\Xi  +\frac{1}{2} [\Lambda N]_{1\frac{1}{2}}\Xi 
                    -\frac{1}{\sqrt{12}} [\Xi N]_{00}\Lambda -\frac{1}{2} [\Xi N]_{10}\Lambda    $ \\
   &  &&& $\Xi \Lambda N\,{v=2}$ &&& $\frac{1}{\sqrt{3}}[\Xi \Lambda]_{1\frac{1}{2}}N -\frac{1}{2} [\Lambda N]_{0\frac{1}{2}}\Xi
                     +\frac{1}{\sqrt{12}} [\Lambda N]_{1\frac{1}{2}}\Xi 
                    -\frac{1}{2} [\Xi N]_{00}\Lambda +\frac{1}{\sqrt{12}} [\Xi N]_{10}\Lambda    $ \\
   &   &&& $\Xi \Sigma N\,{v=1}$ &&& $\frac{1}{\sqrt{3}}[\Xi \Sigma]_{0\frac{1}{2}}N +\frac{1}{\sqrt{12}}
                   [\Sigma N]_{0\frac{1}{2}}\Xi  +\frac{1}{2} [\Sigma N]_{1\frac{1}{2}}\Xi 
                    +\frac{1}{\sqrt{12}} [\Xi N]_{01}\Sigma +\frac{1}{2} [\Xi N]_{11}\Sigma    $ \\
   &  &&& $\Xi \Sigma N\,{v=2}$ &&& $\frac{1}{\sqrt{3}}[\Xi \Sigma]_{1\frac{1}{2}}N -\frac{1}{2} [\Sigma N]_{0\frac{1}{2}}\Xi 
                     +\frac{1}{\sqrt{12}} [\Sigma N]_{1\frac{1}{2}}\Xi 
                    +\frac{1}{2} [\Xi N]_{01}\Sigma -\frac{1}{\sqrt{12}} [\Xi N]_{11}\Sigma    $ \\
  &    &&& $\Sigma \Sigma \Lambda$ &&& $\frac{1}{\sqrt{3}}[\Sigma \Sigma ]_{00}\Lambda -
             \frac{1}{\sqrt{6}}[\Sigma \Lambda]_{01}\Sigma -\frac{1}{\sqrt{2}}[\Sigma \Lambda ]_{11}\Sigma $ \\ 
\hline
 0 & 1 &&& $\Xi \Lambda N\,{v=1}$ &&& $\frac{1}{\sqrt{3}}[\Xi \Lambda]_{0\frac{1}{2}}N -\frac{1}{\sqrt{12}}
                   [\Lambda N]_{0\frac{1}{2}}\Xi  -\frac{1}{2} [\Lambda N]_{1\frac{1}{2}}\Xi 
                    -\frac{1}{\sqrt{12}} [\Xi N]_{01}\Lambda -\frac{1}{2} [\Xi N]_{11}\Lambda    $ \\
   &  &&& $\Xi \Lambda N\,{v=2}$ &&& $\frac{1}{\sqrt{3}}[\Xi \Lambda]_{1\frac{1}{2}}N +\frac{1}{2} [\Lambda N]_{0\frac{1}{2}}\Xi  -\frac{1}{\sqrt{12}} 
                   [\Lambda N]_{1\frac{1}{2}}\Xi -\frac{1}{2} [\Xi N]_{01}\Lambda +\frac{1}{\sqrt{12}} [\Xi N]_{11}\Lambda    $ \\
   &   &&& $\Xi \Sigma N\,{v=1}$ &&& $\frac{1}{\sqrt{3}}[\Xi \Sigma]_{0\frac{1}{2}}N +\frac{1}{\sqrt{108}}
                   [\Sigma N]_{0\frac{1}{2}}\Xi +\sqrt{\frac{2}{27}} [\Sigma N]_{0\frac{3}{2}}\Xi 
                  + \frac{1}{6} [\Sigma N]_{1\frac{1}{2}}\Xi +\frac{\sqrt{2}}{3}[\Sigma N]_{1\frac{3}{2}}\Xi $\\
  &   &&&   &&& $\qquad  -\frac{1}{6}[\Xi N]_{00}\Sigma -\frac{1}{\sqrt{18}}[\Xi N]_{01}\Sigma -\frac{1}{\sqrt{12}} 
                    [\Xi N]_{10}\Sigma -\frac{1}{\sqrt{6}}[\Xi N]_{11}\Sigma   $ \\                   
  &   &&& $\Xi \Sigma N\,{v=2}$ &&& $\frac{1}{\sqrt{3}}[\Xi \Sigma]_{0\frac{3}{2}}N -\sqrt{\frac{2}{27}}
                   [\Sigma N]_{0\frac{1}{2}}\Xi +\frac{1}{\sqrt{108}} [\Sigma N]_{0\frac{3}{2}}\Xi 
                   - \frac{\sqrt{2}}{3} [\Sigma N]_{1\frac{1}{2}}\Xi +\frac{1}{6}[\Sigma N]_{1\frac{3}{2}}\Xi $\\
  &   &&&   &&& $\qquad -\frac{1}{\sqrt{18}}[\Xi N]_{00}\Sigma +\frac{1}{6}[\Xi N]_{01}\Sigma -\frac{1}{\sqrt{6}} 
                    [\Xi N]_{10}\Sigma + \frac{1}{\sqrt{12}}[\Xi N]_{11}\Sigma  $ \\
   &   &&& $\Xi \Sigma N\,{v=3}$ &&& $\frac{1}{\sqrt{3}}[\Xi \Sigma]_{1\frac{1}{2}}N -\frac{1}{6}
                   [\Sigma N]_{0\frac{1}{2}}\Xi -\frac{\sqrt{2}}{3} [\Sigma N]_{0\frac{3}{2}}\Xi 
                   + \frac{1}{\sqrt{108}} [\Sigma N]_{1\frac{1}{2}}\Xi +\sqrt{\frac{2}{27}}[\Sigma N]_{1\frac{3}{2}}\Xi  $\\
  &   &&&   &&& $\qquad  -\frac{1}{\sqrt{12}}[\Xi N]_{00}\Sigma -\frac{1}{\sqrt{6}}[\Xi N]_{01}\Sigma +\frac{1}{6} 
                    [\Xi N]_{10}\Sigma +\frac{1}{\sqrt{18}}[\Xi N]_{11}\Sigma   $ \\
  &   &&& $\Xi \Sigma N\,{v=4}$ &&& $\frac{1}{\sqrt{3}}[\Xi \Sigma]_{1\frac{3}{2}}N +\frac{\sqrt{2}}{3}
                   [\Sigma N]_{0\frac{1}{2}}\Xi -\frac{1}{6} [\Sigma N]_{0\frac{3}{2}}\Xi 
                   - \sqrt{\frac{2}{27}} [\Sigma N]_{1\frac{1}{2}}\Xi +\frac{1}{\sqrt{108}}[\Sigma N]_{1\frac{3}{2}}\Xi $\\
  &   &&&   &&& $\qquad -\frac{1}{\sqrt{6}}[\Xi N]_{00}\Sigma +\frac{1}{\sqrt{12}}[\Xi N]_{01}\Sigma +\frac{1}{\sqrt{18}} 
                    [\Xi N]_{10}\Sigma - \frac{1}{6}[\Xi N]_{11}\Sigma  $ \\                  
  &   &&& $\Lambda \Lambda \Sigma$ &&& $\frac{1}{\sqrt{3}}[\Lambda \Lambda ]_{00}\Sigma -
          \frac{1}{\sqrt{6}}[\Sigma \Lambda]_{01}\Lambda +\frac{1}{\sqrt{2}}[\Sigma \Lambda ]_{11}\Lambda $ \\ 
  &  &&& $\Sigma \Sigma \Lambda$ &&& $\frac{1}{\sqrt{3}}[\Sigma \Sigma ]_{11}\Lambda -
           \frac{1}{\sqrt{2}}[\Sigma \Lambda]_{01}\Sigma +\frac{1}{\sqrt{6}}[\Sigma \Lambda ]_{11}\Sigma $ \\ 
  &  &&& $\Sigma \Sigma \Sigma$ &&& $\frac{\sqrt{2}}{3}[\Sigma\Sigma]_{00}\Sigma 
        -\sqrt{\frac{5}{18}}[\Sigma\Sigma]_{02}\Sigma -\frac{1}{\sqrt{2}}[\Sigma\Sigma]_{11}\Sigma $ \\
\hline
 0 & 2 &&& $\Xi \Sigma N\,{v=1}$ &&& $\frac{1}{\sqrt{3}}[\Xi \Sigma]_{0\frac{3}{2}}N -\frac{1}{\sqrt{12}}
                   [\Sigma N]_{0\frac{3}{2}}\Xi -\frac{1}{2}[\Sigma N]_{1\frac{3}{2}}\Xi 
                 -\frac{1}{\sqrt{12}} [\Xi N]_{01}\Sigma -\frac{1}{2}[\Xi N]_{11}\Sigma $\\  
  &  &&& $\Xi \Sigma N\,{v=2}$ &&& $\frac{1}{\sqrt{3}}[\Xi \Sigma]_{1\frac{3}{2}}N +\frac{1}{2}  [\Sigma N]_{0\frac{3}{2}}\Xi 
                -\frac{1}{\sqrt{12}}[\Sigma N]_{1\frac{3}{2}}\Xi -\frac{1}{2} [\Xi N]_{01}\Sigma +\frac{1}{\sqrt{12}}[\Xi N]_{11}\Sigma $\\  
  &  &&& $\Sigma \Sigma \Lambda$ &&& $\frac{1}{\sqrt{3}}[\Sigma \Sigma ]_{02}\Lambda -
           \frac{1}{\sqrt{6}}[\Sigma \Lambda]_{01}\Sigma -\frac{1}{\sqrt{2}}[\Sigma \Lambda ]_{11}\Sigma $ \\ 
  &  &&& $\Sigma \Sigma \Sigma$ &&& $\frac{1}{\sqrt{2}}[\Sigma\Sigma]_{02}\Sigma -\frac{1}{\sqrt{2}}[\Sigma\Sigma]_{11}\Sigma $ \\
%
$-1$  & $\frac{1}{2}$ &&& $\Xi \Xi N\, {v=1}$ &&& $\frac{1}{\sqrt{3}}[\Xi \Xi]_{01}N -\frac{1}{\sqrt{8}}[\Xi N]_{00}\Xi
               +\frac{1}{\sqrt{24}}[\Xi N]_{01}\Xi -\sqrt{\frac{3}{8}}[\Xi N]_{10}\Xi +\frac{1}{\sqrt{8}}[\Xi N]_{11}\Xi $ \\
              &  &&& $\Xi \Xi N\, {v=2}$ &&& $\frac{1}{\sqrt{3}}[\Xi \Xi]_{10}N -\frac{1}{\sqrt{8}}[\Xi N]_{00}\Xi
            -\sqrt{\frac{3}{8}}[\Xi N]_{01}\Xi +\frac{1}{\sqrt{24}}[\Xi N]_{10}\Xi +\frac{1}{\sqrt{8}}[\Xi N]_{11}\Xi $ \\
          &  &&& $\Xi \Lambda \Lambda$ &&& $\frac{1}{\sqrt{3}}[\Lambda \Lambda ]_{00}\Xi -
           \frac{1}{\sqrt{6}}[\Xi \Lambda]_{0\frac{1}{2}}\Lambda +\frac{1}{\sqrt{2}}[\Xi \Lambda ]_{1\frac{1}{2}}\Lambda $ \\ 
          &  &&& $\Xi \Sigma \Sigma\,{v=1}$ &&& $\frac{1}{\sqrt{3}}[\Sigma \Sigma]_{00}\Xi +\frac{1}{\sqrt{18}}
                [\Xi \Sigma]_{0\frac{1}{2}}\Sigma -\frac{1}{3} [\Xi \Sigma]_{0\frac{3}{2}}\Sigma -\frac{1}{\sqrt{6}} 
                [\Xi \Sigma]_{1\frac{1}{2}}\Sigma +\frac{1}{\sqrt{3}} [\Xi \Sigma]_{1\frac{3}{2}}\Sigma   $ \\
          &  &&& $\Xi \Sigma \Sigma\,{v=2}$ &&& $\frac{1}{\sqrt{3}}[\Sigma \Sigma]_{11}\Xi +\frac{1}{\sqrt{3}}
                [\Xi \Sigma]_{0\frac{1}{2}}\Sigma +\frac{1}{\sqrt{6}} [\Xi \Sigma]_{0\frac{3}{2}}\Sigma +\frac{1}{3} 
                [\Xi \Sigma]_{1\frac{1}{2}}\Sigma +\frac{1}{\sqrt{18}} [\Xi \Sigma]_{1\frac{3}{2}}\Sigma   $ \\
  &  &&& $\Xi \Sigma \Lambda\,{v=1}$ &&&  $\frac{1}{\sqrt{3}}[\Xi \Sigma]_{0\frac{1}{2}}\Lambda +\frac{1}{\sqrt{12}}
                   [\Sigma \Lambda]_{01}\Xi +\frac{1}{2}[\Sigma \Lambda]_{11}\Xi 
                -\frac{1}{\sqrt{12}} [\Xi \Lambda]_{0\frac{1}{2}}\Sigma -\frac{1}{2}[\Xi \Lambda]_{1\frac{1}{2}}\Sigma $ \\   
  &  &&& $\Xi \Sigma \Lambda\,{v=2}$ &&&  $\frac{1}{\sqrt{3}}[\Xi \Sigma]_{1\frac{1}{2}}\Lambda -\frac{1}{2}[\Sigma \Lambda]_{01}\Xi 
               +\frac{1}{\sqrt{12}}[\Sigma \Lambda]_{11}\Xi -\frac{1}{2} [\Xi \Lambda]_{0\frac{1}{2}}\Sigma  
               +\frac{1}{\sqrt{12}}[\Xi \Lambda]_{1\frac{1}{2}}\Sigma $ \\       
\hline
$-1$  & $\frac{3}{2}$ &&& $\Xi \Xi N$ &&& $\frac{1}{\sqrt{3}}[\Xi \Xi]_{01}N -\frac{1}{\sqrt{6}}[\Xi N]_{01}\Xi
                -\frac{1}{\sqrt{2}}[\Xi N]_{11}\Xi $ \\
     &   &&& $\Xi \Sigma \Sigma\,{v=1}$ &&& $\frac{1}{\sqrt{3}}[\Sigma \Sigma]_{02}\Xi +\frac{\sqrt{5}}{6}
                [\Xi \Sigma]_{0\frac{1}{2}}\Sigma +\frac{1}{6} [\Xi \Sigma]_{0\frac{3}{2}}\Sigma -\sqrt{\frac{5}{12}} 
                [\Xi \Sigma]_{1\frac{1}{2}}\Sigma -\frac{1}{\sqrt{12}} [\Xi \Sigma]_{1\frac{3}{2}}\Sigma   $ \\
     &   &&&  $\Xi \Sigma \Sigma\,{v=2}$  &&& $\frac{1}{\sqrt{3}}[\Sigma \Sigma]_{11}\Xi +\frac{1}{\sqrt{12}}
                [\Xi \Sigma]_{0\frac{1}{2}}\Sigma - \sqrt{\frac{5}{12}} [\Xi \Sigma]_{0\frac{3}{2}}\Sigma +\frac{1}{6} 
                [\Xi \Sigma]_{1\frac{1}{2}}\Sigma -\frac{\sqrt{5}}{6} [\Xi \Sigma]_{1\frac{3}{2}}\Sigma   $ \\
  &  &&& $\Xi \Sigma \Lambda\,{v=1}$ &&&  $\frac{1}{\sqrt{3}}[\Xi \Sigma]_{0\frac{3}{2}}\Lambda -\frac{1}{\sqrt{12}}
                   [\Sigma \Lambda]_{01}\Xi  -\frac{1}{2}[\Sigma \Lambda]_{11}\Xi 
                -\frac{1}{\sqrt{12}} [\Xi \Lambda]_{0\frac{1}{2}}\Sigma -\frac{1}{2}[\Xi \Lambda]_{1\frac{1}{2}}\Sigma $ \\   
  &  &&& $\Xi \Sigma \Lambda\,{v=2}$ &&&  $\frac{1}{\sqrt{3}}[\Xi \Sigma]_{1\frac{3}{2}}\Lambda +\frac{1}{2}[\Sigma \Lambda]_{01}\Xi 
               -\frac{1}{\sqrt{12}}[\Sigma \Lambda]_{11}\Xi -\frac{1}{2} [\Xi \Lambda]_{0\frac{1}{2}}\Sigma 
               +\frac{1}{\sqrt{12}}[\Xi \Lambda]_{1\frac{1}{2}}\Sigma $ \\      
\hline
$-1$ & $\frac{5}{2}$ &&& $\Xi \Sigma \Sigma$ &&& $\frac{1}{\sqrt{3}}[\Sigma \Sigma]_{02}\Xi -\frac{1}{\sqrt{6}}
                [\Xi \Sigma]_{0\frac{3}{2}}\Sigma + \frac{1}{\sqrt{2}}[\Xi \Sigma]_{1\frac{3}{2}}\Sigma $ \\
\hline
 $-2$ & 0 &&& $\Xi \Xi \Lambda$ &&& $\frac{1}{\sqrt{3}}[\Xi \Xi]_{10}\Lambda -\frac{1}{\sqrt{2}}
               [\Xi \Lambda]_{0\frac{1}{2}}\Xi +\frac{1}{\sqrt{6}}[\Xi \Lambda]_{1\frac{1}{2}}\Xi $ \\ 
   &  &&& $\Xi \Xi \Sigma$ &&& $\frac{1}{\sqrt{3}}[\Xi \Xi]_{01}\Sigma +\frac{1}{\sqrt{6}}
              [\Xi \Sigma]_{0\frac{1}{2}}\Xi  +\frac{1}{\sqrt{2}}[\Xi \Sigma]_{1\frac{1}{2}}\Xi $ \\ 
\hline
 $-2$ & 1 &&& $\Xi \Xi \Lambda$ &&& $\frac{1}{\sqrt{3}}[\Xi \Xi]_{01}\Lambda -\frac{1}{\sqrt{6}}
             [\Xi \Lambda]_{0\frac{1}{2}}\Xi -\frac{1}{\sqrt{2}}[\Xi \Lambda]_{1\frac{1}{2}}\Xi $ \\ 
   &  &&& $\Xi \Xi \Sigma\,{v=1}$ &&& $\frac{1}{\sqrt{3}}[\Xi \Xi]_{01}\Sigma -\frac{1}{3}
             [\Xi \Sigma]_{0\frac{1}{2}}\Xi +\frac{1}{\sqrt{18}}[\Xi \Sigma]_{0\frac{3}{2}}\Xi -\frac{1}{\sqrt{3}}
            [\Xi \Sigma]_{1\frac{1}{2}}\Xi +\frac{1}{\sqrt{6}}[\Xi \Sigma]_{1\frac{3}{2}}\Xi $ \\
   &  &&& $\Xi \Xi \Sigma\,{v=2}$ &&& $\frac{1}{\sqrt{3}}[\Xi \Xi]_{10}\Sigma -\frac{1}{\sqrt{6}}[\Xi \Sigma]_{0\frac{1}{2}}\Xi 
          -\frac{1}{\sqrt{3}}[\Xi \Sigma]_{0\frac{3}{2}}\Xi +\frac{1}{\sqrt{18}}[\Xi \Sigma]_{1\frac{1}{2}}\Xi 
           +\frac{1}{3}[\Xi \Sigma]_{1\frac{3}{2}}\Xi $ \\
\hline
 $-2$  & 2 &&& $\Xi \Xi \Sigma$ &&& $\frac{1}{\sqrt{3}}[\Xi \Xi]_{01}\Sigma -\frac{1}{\sqrt{6}}
           [\Xi \Sigma]_{0\frac{3}{2}}\Xi   -\frac{1}{\sqrt{2}}[\Xi \Sigma]_{1\frac{3}{2}}\Xi $ \\ 
\hline            
 $-3$ & $\frac{1}{2}$ &&& $\Xi \Xi \Xi$ &&& $\frac{1}{\sqrt{2}}[\Xi \Xi]_{01}\Xi -\frac{1}{\sqrt{2}}[\Xi \Xi]_{10}\Xi $ \\
\end{tabular}
\end{ruledtabular}
\end{table*}

\begin{table*}[h]
\caption{Same as Table~I but for $S=3/2$.}
\label{B_8B_8B_8spin-isospinS=3/2}
\begin{ruledtabular}
\begin{tabular}{cccclcclc}
$Y$ & $I$ &&& $B_8B_8B_8$ &&&  \\
\hline
 2 & 0 &&& $\Lambda NN$ &&& $\frac{1}{\sqrt{3}}[NN]_{10}\Lambda +\sqrt{\frac{2}{3}}[\Lambda N]_{1 \frac{1}{2}}N $ \\ 
\hline
 2  & 1  &&& $\Sigma NN$ &&& $\frac{1}{\sqrt{3}}[NN]_{10}\Sigma -\frac{\sqrt{2}}{3}[\Sigma N]_{1 \frac{1}{2}}N 
                +\frac{2}{3}[\Sigma N]_{1\frac{3}{2}}N $ \\ 
\hline
1 & $\frac{1}{2}$  &&&$\Xi NN$ &&& $\frac{1}{\sqrt{3}}[NN]_{10}\Xi -\frac{1}{\sqrt{6}}[\Xi N]_{10}N 
             +\frac{1}{\sqrt{2}} [\Xi N]_{11}N $ \\
         &   &&& $\Sigma \Sigma N$ &&& $\frac{1}{\sqrt{3}}[\Sigma \Sigma]_{11}N -\frac{2}{3}
             [\Sigma N]_{1\frac{1}{2}}\Sigma +\frac{\sqrt{2}}{3} [\Sigma N]_{1\frac{3}{2}}\Sigma $ \\ 
     &  &&& $\Sigma \Lambda N$ &&& $\frac{1}{\sqrt{3}}[\Sigma \Lambda]_{11}N -\frac{1}{\sqrt{3}}[\Lambda N]_{1\frac{1}{2}}\Sigma 
             -\frac{1}{\sqrt{3}} [\Sigma N]_{1\frac{1}{2}}\Lambda  $ \\
\hline
1 & $\frac{3}{2}$ &&& $\Sigma \Sigma N$ &&& $\frac{1}{\sqrt{3}}[\Sigma \Sigma]_{11}N -\frac{1}{3}
  [\Sigma N]_{1\frac{1}{2}}\Sigma -\frac{\sqrt{5}}{3} [\Sigma N]_{1\frac{3}{2}}\Sigma $ \\
     &  &&& $\Sigma \Lambda N$ &&& $\frac{1}{\sqrt{3}}[\Sigma \Lambda]_{11}N +\frac{1}{\sqrt{3}}[\Lambda N]_{1\frac{1}{2}}\Sigma 
             -\frac{1}{\sqrt{3}} [\Sigma N]_{1\frac{3}{2}}\Lambda  $ \\
\hline
0 & 0 &&& $\Xi \Lambda N$ &&& $\frac{1}{\sqrt{3}}[\Xi \Lambda]_{1\frac{1}{2}}N -\frac{1}{\sqrt{3}}[\Lambda N]_{1\frac{1}{2}}\Xi 
             -\frac{1}{\sqrt{3}} [\Xi N]_{10}\Lambda  $ \\
  &  &&&  $\Xi \Sigma N$ &&& $\frac{1}{\sqrt{3}}[\Xi \Sigma]_{1\frac{1}{2}}N -\frac{1}{\sqrt{3}}[\Sigma N]_{1\frac{1}{2}}\Xi 
             +\frac{1}{\sqrt{3}} [\Xi N]_{11}\Sigma  $ \\
 &  &&& $\Sigma \Sigma \Sigma$ &&& $[\Sigma\Sigma]_{11}\Sigma $ \\
\hline
0 & 1 &&& $\Xi \Lambda N$ &&& $\frac{1}{\sqrt{3}}[\Xi \Lambda]_{1\frac{1}{2}}N +\frac{1}{\sqrt{3}}[\Lambda N]_{1\frac{1}{2}}\Xi 
             -\frac{1}{\sqrt{3}} [\Xi N]_{11}\Lambda  $ \\
  &  &&&  $\Xi \Sigma N\,{v=1}$ &&& $\frac{1}{\sqrt{3}}[\Xi \Sigma]_{1\frac{1}{2}}N -\frac{1}{\sqrt{27}}[\Sigma N]_{1\frac{1}{2}}\Xi 
         -\sqrt{\frac{8}{27}}[\Sigma N]_{1\frac{3}{2}}\Xi -\frac{1}{3} [\Xi N]_{10}\Sigma -\frac{\sqrt{2}}{3} [\Xi N]_{11}\Sigma $ \\
  &  &&& $\Xi \Sigma N\,{v=2}$  &&& $\frac{1}{\sqrt{3}}[\Xi \Sigma]_{1\frac{3}{2}}N +\sqrt{\frac{8}{27}}[\Sigma N]_{1\frac{1}{2}}\Xi 
         -\frac{1}{\sqrt{27}}[\Sigma N]_{1\frac{3}{2}}\Xi -\frac{\sqrt{2}}{3} [\Xi N]_{10}\Sigma +\frac{1}{3} [\Xi N]_{11}\Sigma $ \\          
 &   &&& $\Sigma \Sigma \Lambda$ &&& $\frac{1}{\sqrt{3}}[\Sigma \Sigma ]_{11}\Lambda -
          \sqrt{\frac{2}{3}}[\Sigma \Lambda]_{11}\Sigma $ \\
\hline
0  & 2 &&&  $\Xi \Sigma N$ &&& $\frac{1}{\sqrt{3}}[\Xi \Sigma]_{1\frac{3}{2}}N +\frac{1}{\sqrt{3}}[\Sigma N]_{1\frac{3}{2}}\Xi 
             -\frac{1}{\sqrt{3}} [\Xi N]_{11}\Sigma  $ \\
\hline
$-1$ & $\frac{1}{2}$  &&& $\Xi \Xi N$ &&& $\frac{1}{\sqrt{3}}[\Xi \Xi]_{10}N -\frac{1}{\sqrt{6}}
       [\Xi N]_{10}\Xi -\frac{1}{\sqrt{2}}[\Xi N]_{11}\Xi $ \\
   &   &&& $\Xi \Sigma \Sigma$ &&& $\frac{1}{\sqrt{3}}[\Sigma \Sigma]_{11}\Xi -\frac{2}{3}
       [\Xi \Sigma]_{1\frac{1}{2}}\Sigma -\frac{\sqrt{2}}{3}[\Xi \Sigma]_{1\frac{3}{2}}\Sigma $ \\ 
  &   &&& $\Xi \Sigma \Lambda$ &&& $\frac{1}{\sqrt{3}}[\Xi \Sigma]_{1\frac{1}{2}}\Lambda -\frac{1}{\sqrt{3}}[\Sigma \Lambda]_{11}\Xi 
             -\frac{1}{\sqrt{3}} [\Xi \Lambda]_{1\frac{1}{2}}\Sigma  $ \\ 
\hline
 $-1$  & $\frac{3}{2}$  &&& $\Xi \Sigma \Sigma$ &&& $\frac{1}{\sqrt{3}}[\Sigma \Sigma]_{11}\Xi -\frac{1}{3}
       [\Xi \Sigma]_{1\frac{1}{2}}\Sigma +\frac{\sqrt{5}}{3}[\Xi \Sigma]_{1\frac{3}{2}}\Sigma $ \\ 
  &   &&& $\Xi \Sigma \Lambda$ &&& $\frac{1}{\sqrt{3}}[\Xi \Sigma]_{1\frac{3}{2}}\Lambda +\frac{1}{\sqrt{3}}[\Sigma \Lambda]_{11}\Xi 
             -\frac{1}{\sqrt{3}} [\Xi \Lambda]_{1\frac{1}{2}}\Sigma  $ \\ 
\hline
$-2$ & 0 &&& $\Xi \Xi \Lambda$ &&& $\frac{1}{\sqrt{3}}[\Xi \Xi]_{10}\Lambda -\sqrt{\frac{2}{3}}
         [\Xi \Lambda ]_{1\frac{1}{2}}\Xi $ \\  
\hline
$-2$ & 1 &&& $\Xi \Xi \Sigma$ &&& $\frac{1}{\sqrt{3}}[\Xi \Xi]_{10}\Sigma -\frac{\sqrt{2}}{3}
        [\Xi \Sigma]_{1\frac{1}{2}}\Xi -\frac{2}{3}[\Xi \Sigma]_{1\frac{3}{2}}\Xi $ \\         
\end{tabular}
\end{ruledtabular}
\end{table*}

\newpage
~
\newpage
~

\section{Coefficeints $G$ in Eq.~(12) of the main text}

\begin{table*}[h]
\caption{\label{su3_half}
Coefficients $G$ for some three-$B_8$ systems with $S=1/2$. This table continues.
}
\begin{ruledtabular}
\begin{tabular}{cccccccccc}
$YI$ & \multicolumn{9}{c}{01} \\
\cline{1-1}\cline{2-10}
 & $\Xi\Lambda N$ & $\Xi\Lambda N$ & $\Xi\Sigma N$ & $\Xi\Sigma N$ & $\Xi\Sigma N$
 & $\Xi\Sigma N$ & $\Sigma\Lambda\Lambda$ & $\Sigma\Sigma\Lambda$ & $\Sigma\Sigma\Sigma$ \\ 
 & v=1 & v=2 & v=1 & v=2 & v=3 & v=4 & & & \\
\cline{1-1}\cline{2-10}
$|41\rangle_1$ & $\frac{1}{2\sqrt{6}}$ & $-\frac{1}{2\sqrt{2}}$ & $-\frac{1}{6\sqrt{6}}$ & $-\frac{1}{6\sqrt{3}}$
 & $-\frac{1}{2\sqrt{2}}$ &  & $-\frac{1}{2\sqrt{2}}$ & $-\frac{1}{4}$ & $\frac{1}{12}$  \\
$|41\rangle_2$ & $-\frac{1}{2\sqrt{6}}$ & $\frac{1}{2\sqrt{2}}$ & $\frac{1}{6\sqrt{6}}$ & $\frac{1}{6\sqrt{3}}$
 & $\frac{1}{2\sqrt{2}}$ &  & $\frac{1}{2\sqrt{2}}$ & $\frac{1}{4}$ & $-\frac{1}{12}$  \\
$|30\rangle_1$ &  & $-\frac{1}{\sqrt{10}}$ & $\frac{1}{\sqrt{30}}$ & $\frac{1}{\sqrt{15}}$
 &  &  &  & $\frac{1}{2\sqrt{5}}$ & $-\frac{1}{2\sqrt{5}}$  \\
$|30\rangle_2$ &  & $\frac{1}{3\sqrt{2}}$ & $-\frac{1}{3\sqrt{6}}$ & $-\frac{1}{3\sqrt{3}}$ &  &  &  & $-\frac{1}{6}$ & $\frac{1}{6}$  \\
$|30\rangle_3$ &  & $-\frac{1}{3}$ & $\frac{1}{3\sqrt{3}}$ & $\frac{1}{3}\sqrt{\frac{2}{3}}$ &  &  & 
 & $\frac{1}{3\sqrt{2}}$ & $-\frac{1}{3\sqrt{2}}$  \\
$|30\rangle_4$ &  & $-\frac{1}{\sqrt{15}}$ & $\frac{1}{3\sqrt{5}}$ & $\frac{1}{3}\sqrt{\frac{2}{5}}$ &  &  & 
 & $\frac{1}{\sqrt{30}}$ & $-\frac{1}{\sqrt{30}}$  \\
$|22\rangle_1$ & $-\frac{3}{4\sqrt{5}}$ & $\frac{1}{4}\sqrt{\frac{3}{5}}$ & $\frac{1}{4\sqrt{5}}$ & $\frac{1}{4}\sqrt{\frac{5}{2}}$
 & $-\frac{1}{4}\sqrt{\frac{5}{3}}$ & $-\frac{1}{4\sqrt{30}}$ &  & $-\frac{1}{2}\sqrt{\frac{3}{10}}$ &   \\
$|22\rangle_2$ & $\frac{1}{20}\sqrt{\frac{7}{3}}$ & $\frac{\sqrt{7}}{20}$ & $-\frac{\sqrt{21}}{20}$ & $\frac{1}{20}\sqrt{\frac{21}{2}}$
 & $\frac{\sqrt{7}}{20}$ & $-\frac{1}{20}\sqrt{\frac{7}{2}}$ & $-\frac{\sqrt{7}}{10}$ &  & $-\frac{1}{10}\sqrt{\frac{7}{2}}$  \\
$|22\rangle_3$ & $\frac{1}{2\sqrt{10}}$ & $\frac{1}{2\sqrt{30}}$ & $-\frac{1}{6}\sqrt{\frac{5}{2}}$ & $\frac{1}{6\sqrt{5}}$
 & $\frac{7}{6\sqrt{30}}$ & $-\frac{1}{3\sqrt{15}}$ & $-\frac{1}{2}\sqrt{\frac{3}{10}}$ & $\frac{1}{4\sqrt{15}}$
 & $-\frac{1}{4}\sqrt{\frac{3}{5}}$  \\
$|22\rangle_4$ &  & $\frac{1}{\sqrt{30}}$ & $-\frac{1}{3}\sqrt{\frac{2}{5}}$ & $\frac{7}{12\sqrt{5}}$
 & $\frac{1}{3\sqrt{30}}$ & $-\frac{1}{12}\sqrt{\frac{5}{3}}$ & $-\frac{1}{2}\sqrt{\frac{3}{10}}$ & $-\frac{1}{4\sqrt{15}}$
 & $-\frac{1}{4}\sqrt{\frac{3}{5}}$  \\
$|22\rangle_5$ & $\frac{1}{\sqrt{10}}$ & $-\frac{1}{\sqrt{30}}$ & $-\frac{1}{3\sqrt{10}}$ & $-\frac{\sqrt{5}}{6}$
 & $\frac{1}{3}\sqrt{\frac{5}{6}}$ & $\frac{1}{6\sqrt{15}}$ &  & $\frac{1}{\sqrt{15}}$ &   \\
$|22\rangle_6$ & $-\frac{1}{5\sqrt{6}}$ & $-\frac{1}{5\sqrt{2}}$ & $\frac{1}{5}\sqrt{\frac{3}{2}}$ & $-\frac{\sqrt{3}}{10}$
 & $-\frac{1}{5\sqrt{2}}$ & $\frac{1}{10}$ & $\frac{\sqrt{2}}{5}$ &  & $\frac{1}{5}$  \\
$|14\rangle_1$ & $\frac{1}{\sqrt{6}}$ &  & $-\frac{1}{3\sqrt{6}}$ & $\frac{5}{12\sqrt{3}}$ &
  & $\frac{1}{4}$ & $\frac{1}{2\sqrt{2}}$ & $-\frac{1}{4}$ & $-\frac{1}{12}$  \\
$|14\rangle_2$ & $-\frac{1}{\sqrt{6}}$ &  & $\frac{1}{3\sqrt{6}}$ & $-\frac{5}{12\sqrt{3}}$
 &  & $-\frac{1}{4}$ & $-\frac{1}{2\sqrt{2}}$ & $\frac{1}{4}$ & $\frac{1}{12}$  \\
$|11\rangle_1$ & $-\frac{1}{5\sqrt{6}}$ & $-\frac{1}{5\sqrt{2}}$ & $-\frac{7}{30\sqrt{6}}$ & $\frac{11}{30\sqrt{3}}$
 & $\frac{3}{10\sqrt{2}}$ & $\frac{1}{10}$ & $-\frac{1}{10\sqrt{2}}$ &  & $\frac{11}{30}$  \\
$|11\rangle_2$ & $\frac{1}{2\sqrt{15}}$ & $\frac{1}{6\sqrt{5}}$ & $\frac{1}{3}\sqrt{\frac{5}{3}}$ & $\frac{1}{3\sqrt{30}}$
 & $\frac{2}{3\sqrt{5}}$ & $\frac{1}{3\sqrt{10}}$ & $-\frac{1}{2\sqrt{5}}$ & $-\frac{1}{3}\sqrt{\frac{2}{5}}$ & $\frac{1}{3\sqrt{10}}$  \\
$|11\rangle_3$ &  & $\frac{1}{3\sqrt{5}}$ & $\frac{1}{6\sqrt{15}}$ & $\frac{1}{3\sqrt{30}}$
 & $-\frac{1}{6\sqrt{5}}$ & $\frac{1}{3}\sqrt{\frac{5}{2}}$ & $-\frac{1}{2\sqrt{5}}$ & $\frac{1}{3}\sqrt{\frac{2}{5}}$
 & $\frac{1}{3\sqrt{10}}$  \\
$|11\rangle_4$ & $\frac{1}{4\sqrt{3}}$ & $\frac{1}{4}$ & $\frac{5}{12\sqrt{3}}$ & $-\frac{1}{3}\sqrt{\frac{2}{3}}$
 & $-\frac{1}{4}$ &  &  &  & $-\frac{\sqrt{2}}{3}$  \\
$|11\rangle_5$ & $\frac{1}{4\sqrt{15}}$ & $-\frac{1}{12\sqrt{5}}$ & $\frac{1}{4}\sqrt{\frac{3}{5}}$ & 
 & $\frac{\sqrt{5}}{12}$ & $-\frac{1}{3}\sqrt{\frac{2}{5}}$ &  & $-\frac{1}{3}\sqrt{\frac{2}{5}}$ &   \\
$|11\rangle_6$ & $-\frac{1}{4\sqrt{5}}$ & $\frac{1}{4\sqrt{15}}$ & $-\frac{3}{4\sqrt{5}}$ & 
 & $-\frac{1}{4}\sqrt{\frac{5}{3}}$ & $\sqrt{\frac{2}{15}}$ &  & $\sqrt{\frac{2}{15}}$ &   \\
$|11\rangle_7$ & $\frac{3}{20}$ & $\frac{3\sqrt{3}}{20}$ & $\frac{23}{60}$ & $-\frac{\sqrt{2}}{15}$
 & $-\frac{1}{20\sqrt{3}}$ & $\frac{2}{5}\sqrt{\frac{2}{3}}$ & $-\frac{2}{5\sqrt{3}}$ &  & $-\frac{1}{5}\sqrt{\frac{2}{3}}$  \\
$|11\rangle_8$ &  &  & $-\frac{1}{6\sqrt{2}}$ & $-\frac{1}{6}$
 & $-\frac{1}{2\sqrt{6}}$ & $-\frac{1}{2\sqrt{3}}$ & $\frac{1}{2\sqrt{6}}$ &  & $-\frac{1}{2\sqrt{3}}$  \\
$|03\rangle_1$ & $\frac{1}{2}\sqrt{\frac{3}{10}}$ & $\frac{1}{2\sqrt{10}}$ & $\frac{1}{2\sqrt{30}}$ & $\frac{1}{2\sqrt{15}}$
 & $-\frac{1}{2\sqrt{10}}$ & $-\frac{1}{2\sqrt{5}}$ &  & $\frac{1}{2\sqrt{5}}$ & $\frac{1}{2\sqrt{5}}$  \\
$|03\rangle_2$ & $-\frac{1}{2\sqrt{6}}$ & $-\frac{1}{6\sqrt{2}}$ & $-\frac{1}{6\sqrt{6}}$ & $-\frac{1}{6\sqrt{3}}$
 & $\frac{1}{6\sqrt{2}}$ & $\frac{1}{6}$ &  & $-\frac{1}{6}$ & $-\frac{1}{6}$  \\
$|03\rangle_3$ & $-\frac{1}{2\sqrt{3}}$ & $-\frac{1}{6}$ & $-\frac{1}{6\sqrt{3}}$ & $-\frac{1}{3\sqrt{6}}$
 & $\frac{1}{6}$ & $\frac{1}{3\sqrt{2}}$ &  & $-\frac{1}{3\sqrt{2}}$ & $-\frac{1}{3\sqrt{2}}$  \\
$|03\rangle_4$ & $\frac{1}{2\sqrt{5}}$ & $\frac{1}{2\sqrt{15}}$ & $\frac{1}{6\sqrt{5}}$ & $\frac{1}{3\sqrt{10}}$
 & $-\frac{1}{2\sqrt{15}}$ & $-\frac{1}{\sqrt{30}}$ &  & $\frac{1}{\sqrt{30}}$ & $\frac{1}{\sqrt{30}}$  \\
&&&&&&&&&{\Large{(contd.)}}
\end{tabular}
\end{ruledtabular}
\end{table*}
\begin{table*}[t]
\begin{ruledtabular}
\begin{tabular}{ccccccccccccc}
$YI$ & \multicolumn{7}{c}{$-1\frac{1}{2}$} & \multicolumn{5}{c}{$-1\frac{3}{2}$} \\
\cline{1-1}\cline{2-8}\cline{9-13}
 & $\Xi\Xi N$ & $\Xi\Xi N$ & $\Xi\Lambda\Lambda$ & $\Xi\Sigma\Sigma$ & $\Xi\Sigma\Sigma$
 & $\Xi\Sigma\Lambda$ & $\Xi\Sigma\Lambda$ & $\Xi\Xi N$ & $\Xi\Sigma\Sigma$
 & $\Xi\Sigma\Sigma$ & $\Xi\Sigma\Lambda$ & $\Xi\Sigma\Lambda$ \\ 
 & v=1 & v=2 & & v=1 & v=2 & v=1 & v=2 & & v=1 & v=2 & v=1 & v=2 \\
\cline{1-1}\cline{2-8}\cline{9-13}
$|41\rangle_1$ & $-\frac{1}{4\sqrt{6}}$ & $-\frac{1}{4}\sqrt{\frac{3}{2}}$ & $-\frac{3}{8}$ & $\frac{1}{8\sqrt{3}}$ & $-\frac{1}{4\sqrt{2}}$
 &  $-\frac{1}{8\sqrt{2}}$ & $-\frac{3}{8}\sqrt{\frac{3}{2}}$ &  $\frac{1}{2\sqrt{3}}$ &  $\frac{1}{4}\sqrt{\frac{5}{3}}$ & $-\frac{1}{4}$
 &  $\frac{1}{2}$ &   \\
$|41\rangle_2$ & $\frac{1}{4\sqrt{6}}$ & $\frac{1}{4}\sqrt{\frac{3}{2}}$ & $\frac{3}{8}$ & $-\frac{1}{8\sqrt{3}}$ & $\frac{1}{4\sqrt{2}}$
 &  $\frac{1}{8\sqrt{2}}$ & $\frac{3}{8}\sqrt{\frac{3}{2}}$ &  $-\frac{1}{2\sqrt{3}}$ &  $-\frac{1}{4}\sqrt{\frac{5}{3}}$ & $\frac{1}{4}$
 &  $-\frac{1}{2}$ &   \\
$|30\rangle_1$ & $\frac{1}{4}\sqrt{\frac{3}{5}}$ & $-\frac{1}{4}\sqrt{\frac{3}{5}}$ & $-\frac{3}{4\sqrt{10}}$ & $-\frac{1}{4}\sqrt{\frac{3}{10}}$
 & $\frac{3}{4\sqrt{5}}$ &  $\frac{3}{8\sqrt{5}}$ & $\frac{1}{8}\sqrt{\frac{3}{5}}$ &   &   &   &   &   \\
$|30\rangle_2$ & $-\frac{1}{4\sqrt{3}}$ & $\frac{1}{4\sqrt{3}}$ & $\frac{1}{4\sqrt{2}}$ & $\frac{1}{4\sqrt{6}}$ & $-\frac{1}{4}$
 &  $-\frac{1}{8}$ & $-\frac{1}{8\sqrt{3}}$ &   &   &   &   &   \\
$|30\rangle_3$ & $\frac{1}{2\sqrt{6}}$ & $-\frac{1}{2\sqrt{6}}$ & $-\frac{1}{4}$ & $-\frac{1}{4\sqrt{3}}$ & $\frac{1}{2\sqrt{2}}$ &
 $\frac{1}{4\sqrt{2}}$ & $\frac{1}{4\sqrt{6}}$ &   &   &   &   &   \\
$|30\rangle_4$ & $\frac{1}{2\sqrt{10}}$ & $-\frac{1}{2\sqrt{10}}$ & $-\frac{1}{4}\sqrt{\frac{3}{5}}$ & $-\frac{1}{4\sqrt{5}}$
 & $\frac{1}{2}\sqrt{\frac{3}{10}}$ &  $\frac{1}{4}\sqrt{\frac{3}{10}}$ & $\frac{1}{4\sqrt{10}}$ &   &   &   &   &   \\
$|22\rangle_1$ & $\frac{\sqrt{15}}{8}$ & $-\frac{1}{8}\sqrt{\frac{3}{5}}$ & $\frac{9}{8\sqrt{10}}$ & $-\frac{1}{8}\sqrt{\frac{3}{10}}$
 & $-\frac{1}{8\sqrt{5}}$ &  $\frac{3}{16\sqrt{5}}$ & $-\frac{7}{16}\sqrt{\frac{3}{5}}$ & $-\frac{1}{4}\sqrt{\frac{3}{2}}$
 & $\frac{1}{8}\sqrt{\frac{15}{2}}$ &  $-\frac{5}{8\sqrt{2}}$ & $-\frac{3}{8\sqrt{2}}$ & $\frac{1}{8}\sqrt{\frac{3}{2}}$ \\
$|22\rangle_2$ & $\frac{\sqrt{7}}{40}$ & $\frac{3\sqrt{7}}{40}$ & $-\frac{1}{40}\sqrt{\frac{21}{2}}$ & $-\frac{1}{8}\sqrt{\frac{7}{2}}$
 & $\frac{\sqrt{21}}{40}$ &  $-\frac{17}{80}\sqrt{\frac{7}{3}}$ & $-\frac{3\sqrt{7}}{80}$ &  $\frac{1}{4}\sqrt{\frac{7}{10}}$
 & $-\frac{1}{8}\sqrt{\frac{7}{2}}$ &  $-\frac{1}{8}\sqrt{\frac{21}{10}}$ &  $-\frac{1}{8}\sqrt{\frac{7}{30}}$
 & $\frac{3}{8}\sqrt{\frac{7}{10}}$ \\
$|22\rangle_3$ & $-\frac{1}{20}\sqrt{\frac{5}{6}}$ & $\frac{1}{4}\sqrt{\frac{5}{6}}$ & $-\frac{3}{8\sqrt{5}}$ & $-\frac{7}{8\sqrt{15}}$
 & $\frac{1}{12}\sqrt{\frac{5}{2}}$ &  $-\frac{9}{8\sqrt{10}}$ & $-\frac{1}{8\sqrt{30}}$ &  $\frac{1}{2\sqrt{3}}$
 & $-\frac{1}{4}\sqrt{\frac{5}{3}}$ &  $-\frac{1}{12}$ &  & $\frac{1}{2\sqrt{3}}$ \\
$|22\rangle_4$ & $\frac{1}{\sqrt{30}}$ & $\frac{1}{\sqrt{30}}$ & & $-\frac{1}{\sqrt{15}}$ & $\frac{1}{3\sqrt{10}}$
 & $-\frac{1}{\sqrt{10}}$ & $-\frac{1}{\sqrt{30}}$ &  $\frac{1}{4\sqrt{3}}$
 & $-\frac{1}{8}\sqrt{\frac{5}{3}}$ &  $-\frac{7}{24}$ & $-\frac{1}{8}$ & $\frac{5}{8\sqrt{3}}$ \\
$|22\rangle_5$ & $-\frac{1}{2}\sqrt{\frac{5}{6}}$ & $\frac{1}{2\sqrt{30}}$ & $-\frac{3}{4\sqrt{5}}$ & $\frac{1}{4\sqrt{15}}$
 & $\frac{1}{6\sqrt{10}}$ &  $-\frac{1}{4\sqrt{10}}$ & $\frac{7}{4\sqrt{30}}$ &  $\frac{1}{2\sqrt{3}}$ &  $-\frac{1}{4}\sqrt{\frac{5}{3}}$
 &  $\frac{5}{12}$ & $\frac{1}{4}$ & $-\frac{1}{4\sqrt{3}}$ \\
$|22\rangle_6$ & $-\frac{1}{10\sqrt{2}}$ & $-\frac{3}{10\sqrt{2}}$ & $\frac{\sqrt{3}}{20}$ & $\frac{1}{4}$ & $-\frac{1}{10}\sqrt{\frac{3}{2}}$
 &  $\frac{17}{20\sqrt{6}}$ & $\frac{3}{20\sqrt{2}}$ & $-\frac{1}{2\sqrt{5}}$ &  $\frac{1}{4}$ & $\frac{1}{4}\sqrt{\frac{3}{5}}$
 &  $\frac{1}{4\sqrt{15}}$ & $-\frac{3}{4\sqrt{5}}$ \\
$|14\rangle_1$ &  &  &  &  &  & 
 &  &  $\frac{1}{4}\sqrt{\frac{5}{3}}$ &  $-\frac{1}{8\sqrt{3}}$ & $-\frac{\sqrt{5}}{8}$ & $-\frac{\sqrt{5}}{8}$ & $-\frac{\sqrt{15}}{8}$ \\
$|14\rangle_2$ &  &  &  &  &  & 
 &  &  $-\frac{1}{4}\sqrt{\frac{5}{3}}$ &  $\frac{1}{8\sqrt{3}}$ & $\frac{\sqrt{5}}{8}$ & $\frac{\sqrt{5}}{8}$ & $\frac{\sqrt{15}}{8}$ \\
$|11\rangle_1$ & $\frac{11}{20\sqrt{3}}$ & $\frac{\sqrt{3}}{20}$ & $-\frac{3}{10\sqrt{2}}$ & $\frac{1}{2\sqrt{6}}$ & $-\frac{1}{5}$
 &  $-\frac{1}{10}$ & $\frac{\sqrt{3}}{10}$ &   &   &   &   &   \\
$|11\rangle_2$ & $\frac{1}{2\sqrt{30}}$ & $\frac{1}{2}\sqrt{\frac{5}{6}}$ & $-\frac{1}{2\sqrt{5}}$ & $\frac{1}{2\sqrt{15}}$ & 
 &  $\frac{1}{\sqrt{10}}$ & $-\frac{1}{\sqrt{30}}$ &   &   &   &   &   \\
$|11\rangle_3$ & $\frac{1}{2\sqrt{30}}$ & $\frac{1}{2\sqrt{30}}$ &  & $\frac{2}{\sqrt{15}}$ & $\frac{1}{\sqrt{10}}$ &  $-\frac{1}{2\sqrt{10}}$
 & $-\frac{1}{2\sqrt{30}}$ &   &   &   &   &   \\
$|11\rangle_4$ & $-\frac{1}{\sqrt{6}}$ &  & $\frac{1}{4}$ & $-\frac{1}{4\sqrt{3}}$ & $\frac{1}{2\sqrt{2}}$ &  $\frac{1}{4\sqrt{2}}$
 & $-\frac{1}{4}\sqrt{\frac{3}{2}}$ &   &   &   &   &   \\
$|11\rangle_5$ &  & $\frac{1}{\sqrt{30}}$ & $-\frac{1}{4\sqrt{5}}$ & $-\frac{1}{4}\sqrt{\frac{3}{5}}$ & $-\frac{1}{2\sqrt{10}}$
 &  $\frac{3}{4\sqrt{10}}$ & $-\frac{1}{4\sqrt{30}}$ &   &   &   &   &   \\
$|11\rangle_6$ &  & $-\frac{1}{\sqrt{10}}$ & $\frac{1}{4}\sqrt{\frac{3}{5}}$ & $\frac{3}{4\sqrt{5}}$ & $\frac{1}{2}\sqrt{\frac{3}{10}}$
 &  $-\frac{3}{4}\sqrt{\frac{3}{10}}$ & $\frac{1}{4\sqrt{10}}$ &   &   &   &   &   \\
$|11\rangle_7$ & $-\frac{1}{5\sqrt{2}}$ & $\frac{\sqrt{2}}{5}$ & $\frac{1}{20\sqrt{3}}$ & $\frac{1}{4}$ & $\frac{3}{10}\sqrt{\frac{3}{2}}$
 &  $\frac{3}{20}\sqrt{\frac{3}{2}}$ & $-\frac{9}{20\sqrt{2}}$ &   &   &  &   &   \\
$|11\rangle_8$ & $-\frac{1}{4}$ & $-\frac{1}{4}$ & $\frac{1}{2\sqrt{6}}$ & $-\frac{1}{2\sqrt{2}}$ &  &  
 &  &   &   &   &   &   \\
$|03\rangle_1$ &  &  &  &  &  & 
 &  &  $\frac{1}{2}\sqrt{\frac{3}{10}}$ & $\frac{1}{4}\sqrt{\frac{3}{2}}$ & $\frac{3}{4\sqrt{10}}$ & $-\frac{3}{4\sqrt{10}}$
 & $\frac{1}{4}\sqrt{\frac{3}{10}}$ \\
$|03\rangle_2$ &  &  &  &  &  & 
 &  &  $-\frac{1}{2\sqrt{6}}$ &  $-\frac{1}{4}\sqrt{\frac{5}{6}}$ & $-\frac{1}{4\sqrt{2}}$ & $\frac{1}{4\sqrt{2}}$
 & $-\frac{1}{4\sqrt{6}}$ \\
$|03\rangle_3$ &  &  &  &  &  &  
 &  &  $-\frac{1}{2\sqrt{3}}$ & $-\frac{1}{4}\sqrt{\frac{5}{3}}$ & $-\frac{1}{4}$ & $\frac{1}{4}$ & $-\frac{1}{4\sqrt{3}}$ \\
$|03\rangle_4$ &  &  &  &  &  &  
 &  &  $\frac{1}{2\sqrt{5}}$ &  $\frac{1}{4}$ &  $\frac{1}{4}\sqrt{\frac{3}{5}}$ & $-\frac{1}{4}\sqrt{\frac{3}{5}}$ & $\frac{1}{4\sqrt{5}}$ \\
&&&&&&&&&&&&{\Large{(contd.)}}
\end{tabular}
\end{ruledtabular}
\end{table*}

\begin{table*}[t]
\begin{ruledtabular}
\begin{tabular}{ccccccccccccc}
$YI$ & \multicolumn{5}{c}{00} & \multicolumn{4}{c}{02} & \multicolumn{3}{c}{$-2$1} \\
\cline{1-1}\cline{2-6}\cline{7-10}\cline{11-13}
 & $\Xi\Lambda N$  & $\Xi\Lambda N$ & $\Xi\Sigma N$ & $\Xi\Sigma N$
 & $\Sigma\Sigma\Lambda$ & $\Xi\Sigma N$ & $\Xi\Sigma N$ & $\Sigma\Sigma\Lambda$
 & $\Sigma\Sigma\Sigma$ & $\Xi\Xi\Lambda$ & $\Xi\Xi\Sigma$ & $\Xi\Xi\Sigma$ \\ 
 & v=1 & v=2 & v=1 & v=2 & & v=1 & v=2 & & & & v=1 & v=2 \\
\cline{1-1}\cline{2-6}\cline{7-10}\cline{11-13}
$|41\rangle_1$ &  &  &  &  &  & $\frac{1}{2}$ &  & $\frac{\sqrt{3}}{4}$ & $\frac{1}{4}$
 & $\frac{\sqrt{3}}{4}$ & $\frac{1}{2\sqrt{2}}$ & $-\frac{\sqrt{3}}{4}$ \\
$|41\rangle_2$ &  &  &  &  &  & $-\frac{1}{2}$ &  & $-\frac{\sqrt{3}}{4}$ & $-\frac{1}{4}$
 & $-\frac{\sqrt{3}}{4}$ & $-\frac{1}{2\sqrt{2}}$ & $\frac{\sqrt{3}}{4}$ \\
$|22\rangle_1$ & $\frac{3}{4}\sqrt{\frac{3}{5}}$ & $-\frac{3}{4\sqrt{5}}$ & $-\frac{1}{4}\sqrt{\frac{3}{5}}$ & $\frac{1}{4\sqrt{5}}$
 &  & $-\frac{1}{4}\sqrt{\frac{3}{2}}$ & $\frac{1}{4\sqrt{2}}$ &  & $\frac{1}{2}\sqrt{\frac{3}{2}}$
 & $-\frac{3}{4\sqrt{2}}$ & $\frac{\sqrt{3}}{4}$ & $-\frac{1}{4\sqrt{2}}$ \\
$|22\rangle_2$ & $\frac{\sqrt{7}}{20}$ & $\frac{\sqrt{21}}{20}$ & $\frac{\sqrt{7}}{20}$ & $\frac{\sqrt{21}}{20}$
 & $-\frac{1}{5}\sqrt{\frac{7}{3}}$ & $\frac{1}{4}\sqrt{\frac{7}{10}}$ & $\frac{1}{4}\sqrt{\frac{21}{10}}$
 & $-\frac{1}{2}\sqrt{\frac{7}{30}}$ &  & $\frac{1}{4}\sqrt{\frac{7}{30}}$ & $\frac{1}{4}\sqrt{\frac{7}{5}}$
 & $\frac{1}{4}\sqrt{\frac{21}{10}}$ \\
$|22\rangle_3$ &  & $\frac{1}{\sqrt{10}}$ & $\frac{1}{\sqrt{30}}$ & $\frac{1}{3}\sqrt{\frac{2}{5}}$
 & $-\frac{1}{\sqrt{10}}$ & $\frac{1}{2\sqrt{3}}$ & $\frac{1}{3}$ & $-\frac{1}{4}$ & $-\frac{1}{4\sqrt{3}}$
 & $\frac{1}{4}$ & $\frac{1}{2\sqrt{6}}$ & $\frac{5}{12}$ \\
$|22\rangle_4$ & $\frac{1}{2}\sqrt{\frac{3}{10}}$ & $\frac{1}{2\sqrt{10}}$ & $\frac{1}{2\sqrt{30}}$ & $\frac{1}{6}\sqrt{\frac{5}{2}}$
 & $-\frac{1}{\sqrt{10}}$ & $\frac{1}{4\sqrt{3}}$ & $\frac{5}{12}$ & $-\frac{1}{4}$ & $\frac{1}{4\sqrt{3}}$
 &  & $\frac{1}{\sqrt{6}}$ & $\frac{1}{3}$ \\
$|22\rangle_5$ & $-\sqrt{\frac{3}{10}}$ & $\frac{1}{\sqrt{10}}$ & $\frac{1}{\sqrt{30}}$ & $-\frac{1}{3\sqrt{10}}$
 &  & $\frac{1}{2\sqrt{3}}$ & $-\frac{1}{6}$ &  & $-\frac{1}{\sqrt{3}}$
 & $\frac{1}{2}$ & $-\frac{1}{\sqrt{6}}$ & $\frac{1}{6}$ \\
$|22\rangle_6$ & $-\frac{1}{5\sqrt{2}}$ & $-\frac{1}{5}\sqrt{\frac{3}{2}}$ & $-\frac{1}{5\sqrt{2}}$ & $-\frac{1}{5}\sqrt{\frac{3}{2}}$
 & $\frac{2}{5}\sqrt{\frac{2}{3}}$ & $-\frac{1}{2\sqrt{5}}$ & $-\frac{1}{2}\sqrt{\frac{3}{5}}$ & $\frac{1}{\sqrt{15}}$
 &  & $-\frac{1}{2\sqrt{15}}$ & $-\frac{1}{\sqrt{10}}$ & $-\frac{1}{2}\sqrt{\frac{3}{5}}$ \\
$|14\rangle_1$ &  &  &  &  &  & $\frac{1}{4}$ & $-\frac{\sqrt{3}}{4}$ & $-\frac{\sqrt{3}}{4}$ & $\frac{1}{4}$ &  &  &  \\
$|14\rangle_2$ &  &  &  &  &  & $-\frac{1}{4}$ & $\frac{\sqrt{3}}{4}$ & $\frac{\sqrt{3}}{4}$ & $-\frac{1}{4}$ &  &  &  \\
$|11\rangle_1$ & $\frac{3}{10\sqrt{2}}$ & $\frac{3}{10}\sqrt{\frac{3}{2}}$ & $-\frac{1}{5\sqrt{2}}$ & $-\frac{1}{5}\sqrt{\frac{3}{2}}$
 & $\frac{1}{10}\sqrt{\frac{3}{2}}$ &  &  &  &  &  &  &  \\
$|11\rangle_2$ &  & $\frac{1}{\sqrt{15}}$ & $-\frac{1}{2\sqrt{5}}$ & $\frac{1}{2}\sqrt{\frac{3}{5}}$
 & $\frac{1}{2}\sqrt{\frac{3}{5}}$ &  &  &  &  &  &  &  \\
$|11\rangle_3$ & $\frac{1}{2\sqrt{5}}$ & $\frac{1}{2\sqrt{15}}$ & $\frac{1}{\sqrt{5}}$ & 
 & $\frac{1}{2}\sqrt{\frac{3}{5}}$ &  &  &  &  &  &  &  \\
$|11\rangle_4$ & $-\frac{1}{4}$ & $-\frac{\sqrt{3}}{4}$ & $\frac{1}{4}$ & $\frac{\sqrt{3}}{4}$ &  &  &  &  &  &  &  &  \\
$|11\rangle_5$ & $-\frac{1}{4\sqrt{5}}$ & $\frac{1}{4\sqrt{15}}$ & $-\frac{3}{4\sqrt{5}}$ & $\frac{1}{4}\sqrt{\frac{3}{5}}$
 &  &  &  &  &  &  &  &  \\
$|11\rangle_6$ & $\frac{1}{4}\sqrt{\frac{3}{5}}$ & $-\frac{1}{4\sqrt{5}}$ & $\frac{3}{4}\sqrt{\frac{3}{5}}$ & $-\frac{3}{4\sqrt{5}}$
 &  &  &  &  &  &  &  &  \\
$|11\rangle_7$ & $-\frac{1}{20\sqrt{3}}$ & $-\frac{1}{20}$ & $\frac{3\sqrt{3}}{20}$ & $\frac{9}{20}$
 & $\frac{2}{5}$ &  &  &  &  &  &  &  \\
$|11\rangle_8$ & $-\frac{1}{2\sqrt{6}}$ & $-\frac{1}{2\sqrt{2}}$ &  &  & $-\frac{1}{2\sqrt{2}}$ &  &  &  &  &  &  &  \\
\end{tabular}
\end{ruledtabular}
\end{table*}

\begin{table*}[t]
\caption{\label{su3_three-half}
Same as Table \ref{su3_half} but for $S=3/2$.
}
\begin{ruledtabular}
\begin{tabular}{cccccccccc}
$YI$ & \multicolumn{3}{c}{00} & \multicolumn{4}{c}{01} & \multicolumn{2}{c}{$-1\frac{3}{2}$} \\
\cline{1-1}\cline{2-4}\cline{5-8}\cline{9-10}
 & $\Xi\Lambda N$ & $\Xi\Sigma N$ & $\Sigma\Sigma\Sigma$ & $\Xi\Lambda N$ & $\Xi\Sigma N$
 & $\Xi\Sigma N$ & $\Sigma\Sigma\Lambda$ & $\Xi\Sigma\Sigma$ & $\Xi\Sigma\Lambda$ \\ 
 & & & & & v=1 & v=2 & & & \\
\cline{1-1}\cline{2-4}\cline{5-8}\cline{9-10}
$|30\rangle_2$ &  &  &  & $\frac{1}{3\sqrt{2}}$ & $\frac{1}{\sqrt{2}}$ &  & $\frac{1}{3}$ &  &   \\
$|30\rangle_3$ &  &  &  & $\frac{1}{6}$ & $\frac{1}{2}$ &  & $\frac{1}{3\sqrt{2}}$ &  &   \\
$|22\rangle_3$ & $\frac{1}{\sqrt{10}}$ & $\frac{1}{3}\sqrt{\frac{5}{2}}$ & $\frac{1}{\sqrt{15}}$ & $\sqrt{\frac{2}{15}}$
 & $\frac{1}{3}\sqrt{\frac{2}{15}}$ & $\frac{2}{3\sqrt{15}}$ & $-\frac{2}{\sqrt{15}}$ & $-\frac{1}{3}$ & $-\frac{1}{\sqrt{3}}$  \\
$|22\rangle_4$ & $-\frac{1}{\sqrt{10}}$ & $-\frac{1}{3}\sqrt{\frac{5}{2}}$ & $-\frac{1}{\sqrt{15}}$ & $-\sqrt{\frac{2}{15}}$
 & $-\frac{1}{3}\sqrt{\frac{2}{15}}$ & $-\frac{2}{3\sqrt{15}}$ & $\frac{2}{\sqrt{15}}$ & $\frac{1}{3}$ & $\frac{1}{\sqrt{3}}$  \\
$|22\rangle_5$ & $-\frac{1}{2\sqrt{10}}$ & $-\frac{1}{6}\sqrt{\frac{5}{2}}$ & $-\frac{1}{2\sqrt{15}}$ & $-\frac{1}{\sqrt{30}}$
 & $-\frac{1}{3\sqrt{30}}$ & $-\frac{1}{3\sqrt{15}}$ & $\frac{1}{\sqrt{15}}$ & $\frac{1}{6}$ & $\frac{1}{2\sqrt{3}}$  \\
$|11\rangle_2$ & $\frac{1}{\sqrt{15}}$ &  & $-\frac{1}{\sqrt{10}}$ & $\frac{2}{3\sqrt{5}}$ & $-\frac{1}{3\sqrt{5}}$
 & $-\frac{1}{3}\sqrt{\frac{2}{5}}$ & $\frac{1}{3\sqrt{10}}$ &  &   \\
$|11\rangle_3$ & $-\frac{1}{\sqrt{15}}$ &  & $\frac{1}{\sqrt{10}}$ & $-\frac{2}{3\sqrt{5}}$ & $\frac{1}{3\sqrt{5}}$
 & $\frac{1}{3}\sqrt{\frac{2}{5}}$ & $-\frac{1}{3\sqrt{10}}$ &  &   \\
$|11\rangle_5$ & $-\frac{2}{\sqrt{15}}$ &  & $\sqrt{\frac{2}{5}}$ & $-\frac{4}{3\sqrt{5}}$ & $\frac{2}{3\sqrt{5}}$
 & $\frac{2}{3}\sqrt{\frac{2}{5}}$ & $-\frac{1}{3}\sqrt{\frac{2}{5}}$ &  &   \\
$|03\rangle_2$ &  &  &  & $\frac{1}{3\sqrt{2}}$ & $-\frac{1}{3\sqrt{2}}$ & $\frac{2}{3}$ & $\frac{1}{3}$ & $-\frac{1}{2}$
 & $-\frac{1}{\sqrt{6}}$  \\
$|03\rangle_3$ &  &  &  & $-\frac{1}{6}$ & $\frac{1}{6}$ & $-\frac{\sqrt{2}}{3}$ & $-\frac{1}{3\sqrt{2}}$ & $\frac{1}{\sqrt{2}}$
 & $\frac{1}{2\sqrt{3}}$  \\
$|00\rangle_1$ & $-\frac{1}{2}\sqrt{\frac{3}{2}}$ & $\frac{1}{2}\sqrt{\frac{3}{2}}$ & $-\frac{1}{2}$ &  &  &  &  &  &   \\
\end{tabular}
\end{ruledtabular}
\end{table*}

\newpage
~
\newpage
~

\section{Matrix elements of quark antisymmetrizer for three octet-baryons}

We give the matrix elements of the antisymmetrizer 
\begin{align}
\Big\langle \Psi^{(\rm orb)}(B_1B_2B_3)|\lambda \mu\rangle_n \Psi^{(\rm color)}(B_1B_2B_3)|{\cal A}'|\Psi^{(\rm orb)}(B_1B_2B_3)|\lambda \mu\rangle_{n'} \Psi^{(\rm color)}(B_1B_2B_3)\Big\rangle,
\label{C1}
\end{align}
where $\Psi^{(\rm orb)}(B_1B_2B_3)$ is the $(0s)^9$ configuration and 
$\Psi^{(\rm color)}(B_1B_2B_3)$ is a product of color-singlet functions, 
$C(123)C(456)C(789)$. They are actually independent of $B_1B_2B_3$. The spin-flavor functions $|\lambda\mu\rangle_n$ are defined in Table I of the main text.
Note that the matrix element vanishes for the different SU(3) labels. 
The matrix elements, Eq.(\ref{C1}), are labeled by $|\lambda \mu\rangle_{n}$ and $|\lambda \mu\rangle_{n'}$. See Table I of the main text for detail of $|\lambda\mu\rangle_{n}$.  Other matrix elements that are not listed here 
all vanish.

Matrix elements of ${\cal A}'$ between $|\lambda\mu\rangle_{n}$'s for 
the total spin $S=1/2$ are as follows: 
\begin{align}
\begin{array}{c|cc}
    & |41\rangle_1 & |41\rangle_2 \\ \hline
|41\rangle_1 & \frac{2}{81} & -\frac{2}{81} \\ 
|41\rangle_2 & -\frac{2}{81} & \frac{2}{81} 
\end{array}
\ \ \ \ \ \ \ \ 
\begin{array}{c|cccc}
   & |22\rangle_1 & |22\rangle_3 & |22\rangle_4 & |22\rangle_5 \\ \hline
|22\rangle_1 & \frac{100}{243} & -\frac{50\sqrt{2}}{729} & \frac{50\sqrt{2}}{729} & -\frac{200\sqrt{2}}{729} \\
|22\rangle_3 & -\frac{50\sqrt{2}}{729} & \frac{50}{2187} & -\frac{50}{2187} & \frac{200}{2187} \\
|22\rangle_4 & \frac{50\sqrt{2}}{729} & -\frac{50}{2187} & \frac{50}{2187} & -\frac{200}{2187} \\
|22\rangle_5 & -\frac{200\sqrt{2}}{729} & \frac{200}{2187} & -\frac{200}{2187} & \frac{800}{2187}
\end{array}
\ \ \ \ \ \ \ \ 
\begin{array}{c|cc}
    & |14\rangle_1 & |14\rangle_2 \\ \hline
|14\rangle_1 & \frac{50}{81} & -\frac{50}{81} \\ 
|14\rangle_2 & -\frac{50}{81} & \frac{50}{81} \notag
\end{array}
\end{align}

\begin{align}
\begin{array}{c|ccccc}
   & |11\rangle_1 & |11\rangle_2 & |11\rangle_3 & |11\rangle_4 & |11\rangle_8 \\ \hline
|11\rangle_1 & \frac{125}{324} & \frac{75}{486}\sqrt{\frac{5}{2}} & \frac{75}{486}\sqrt{\frac{5}{2}}
 & -\frac{25\sqrt{2}}{81} & -\frac{25}{36\sqrt{3}} \\
|11\rangle_2 & \frac{75}{486}\sqrt{\frac{5}{2}} & \frac{25}{162} & \frac{25}{162}
 & -\frac{10\sqrt{5}}{81} & -\frac{5}{18}\sqrt{\frac{5}{6}} \\
|11\rangle_3 & \frac{75}{486}\sqrt{\frac{5}{2}} & \frac{25}{162} & \frac{25}{162}
 & -\frac{10\sqrt{5}}{81} & -\frac{5}{18}\sqrt{\frac{5}{6}} \\
|11\rangle_4 & -\frac{25\sqrt{2}}{81} & -\frac{10\sqrt{5}}{81} & -\frac{10\sqrt{5}}{81}
 & \frac{40}{81} & \frac{5}{9}\sqrt{\frac{2}{3}} \\
|11\rangle_8 & -\frac{25}{36\sqrt{3}} & -\frac{5}{18}\sqrt{\frac{5}{6}} & -\frac{5}{18}\sqrt{\frac{5}{6}}
 & \frac{5}{9}\sqrt{\frac{2}{3}} & \frac{5}{12} \notag
\end{array}
\end{align}
Those for $S=3/2$ are as follows:
\begin{align}
\begin{array}{c|cc}
    & |30\rangle_2 & |30\rangle_3 \\ \hline
|30\rangle_2 & \frac{2}{81} & \frac{\sqrt{2}}{81} \\ 
|30\rangle_3 & \frac{\sqrt{2}}{81} & \frac{1}{81} 
\end{array}
\ \ \ \ \ \ \ \ 
\begin{array}{c|ccc}
   & |22\rangle_3 & |22\rangle_4 & |22\rangle_5  \\ \hline
|22\rangle_3 & \frac{140}{2187} & -\frac{140}{2187} & -\frac{70}{2187}  \\
|22\rangle_4 & -\frac{140}{2187} & \frac{140}{2187} & \frac{70}{2187}  \\
|22\rangle_5 & -\frac{70}{2187} & \frac{70}{2187} & \frac{35}{2187}  
\end{array}
\ \ \ \ \ \ \ \ 
\begin{array}{c|ccc}
   & |11\rangle_2 & |11\rangle_3 & |11\rangle_5  \\ \hline
|11\rangle_2 & \frac{5}{54} & -\frac{5}{54} & -\frac{5}{27}  \\
|11\rangle_3 & -\frac{5}{54} & \frac{5}{54} & \frac{5}{27}  \\
|11\rangle_5 & -\frac{5}{27} & \frac{5}{27} & \frac{10}{27}  \notag
\end{array}
\end{align}

\begin{align}
\begin{array}{c|cc}
    & |03\rangle_2 & |03\rangle_3 \\ \hline
|03\rangle_2 & \frac{50}{81} & -\frac{25\sqrt{2}}{81} \\ 
|03\rangle_3 &  -\frac{25\sqrt{2}}{81} & \frac{25}{81} \notag
\end{array}
\ \ \ \ \ \ 
\begin{array}{c|c}
    & |00\rangle_1 \\ \hline
|00\rangle_1 & \frac{35}{27} \notag
\end{array}
\end{align}

\end{widetext}

\nocite{*}

\bibliography{apssamp}